\documentclass[structabstract]{aa}  
\usepackage{graphicx}
\usepackage{txfonts}
\newcommand{\new}[1]{{#1}}
\newcommand{\newII}[1]{{#1}}

\begin{document}
   \title{Massive open star clusters using the VVV survey\thanks{Based on observations made with NTT telescope at the La Silla Observatory, ESO, under programme ID 087.D-0490A, and with the Clay telescope at the Las Campanas Observatory under programme CN2011A-086. Also based on data from the VVV survey observed under program ID 172.B-2002.}}

   \subtitle{I. Presentation of the data and description of the approach}

   \author{A.-N. Chen\'e\inst{1,2}, J. Borissova\inst{1,3}, J.~R.~A. Clarke\inst{1,4}, C. Bonatto\inst{5}, D.~J. Majaess\inst{6}, C. Moni Bidin\inst{2}, S.~E. Sale\inst{1,7}, F. Mauro\inst{2}, R. Kurtev\inst{1}, G. Baume\inst{8}, C. Feinstein\inst{8}, V.~D. Ivanov\inst{9}, D. Geisler\inst{2}, M. Catelan\inst{3,7}, D. Minniti\inst{3,7,10,11},  P. Lucas\inst{4}, R. de Grijs\inst{12} and  M.~S.~N. Kumar\inst{13}
   }
   \authorrunning{Chen\'e et al.}
   \titlerunning{Study of three young, massive open star clusters using VVV}
   
   \institute{Departamento de F\'isica y Astronom\'ia, Universidad de Valpara\'iso, Av. Gran Breta\~na 1111, Playa Ancha, Casilla 5030, Chile\\
              \email{andrenicolas.chene@gmail.com}
         \and
             Departamento de Astronom\'ia, Universidad de Concepci\'on, Casilla 160-C, Chile
         \and
             The Milky Way Millennium Nucleus, Av. Vicu\~{n}a Mackenna 4860, 782-0436 Macul, Santiago, Chile
         \and
             University of Hertfordshire, Hatfield, AL10 9AB, UK
         \and
             Universidade Federal do Rio Grande do Sul, Departamento de Astronomia CP 15051, RS, Porto Alegre, 91501-970, Brazil
         \and
            Saint MaryÕs University, Halifax, Nova Scotia, Canada
         \and
            Pontificia Universidad Cat\'olica de Chile, Facultad de F\'{i}sica, Departamento de Astronom\'\i a y Astrof\'\i sica, Av. Vicu\~{n}a Mackenna 4860, 782-0436 Macul, Santiago, Chile
         \and
             Facultad de Ciencias Astron\'omicas y Geof\'isicas (UNLP), Instituto de Astrof\'isica de La Plata (CONICET, UNLP), Paseo del Bosque s/n, La Plata, Argentina
         \and
             European Southern Observatory, Ave. Alonso de Cordova 3107, Casilla 19, Chile
         \and
             Vatican Observatory, V00120 Vatican City State, Italy
         \and
             Department of Astrophysical Sciences, Princeton University, Princeton NJ 08544-1001
         \and
             Kavli Institute for Astronomy and Astrophysics, Peking University, Yi He Yuan Lu 5, Hai Dian District, Beijing 100871, China
         \and
             Centro de Astrofisica da Universidade do Porto, Rua das Estrelas, 4150-762 Porto, Portugal
             }

   \date{Received February 17, 2012; accepted June 27, 2012}

 \abstract
{The ESO Public Survey ``VISTA Variables in the V\'ia L\'actea'' (VVV) \new{provides deep multi-epoch infrared observations for an unprecedented 562 sq. degrees of the Galactic bulge, and adjacent regions of the disk.}}
{\new{The VVV observations will foster the construction of a sample of Galactic star clusters with reliable and homogeneously derived physical parameters (e.g., age, distance, and mass, etc.). In this first paper in a series, the methodology employed to establish cluster parameters for the envisioned database are elaborated upon by analysing 4 known young open clusters: Danks\,1, Danks\,2, RCW\,79, and DBS\,132. The analysis offers a first glimpse of the information that can be gleaned from the VVV observations for clusters in the final database.}}
{Wide-field, deep $JHK_s$ VVV observations, combined with new infrared spectroscopy, are employed to constrain fundamental parameters for a subset of clusters.}
{\new{
Results are inferred from VVV near-infrared photometry and numerous low resolution spectra (typically more than 10 per cluster). The high quality of the spectra and the deep wide--field VVV photometry enables us to precisely and independently determine the characteristics of the clusters studied, which we compare to previous determinations. An anomalous reddening law in the direction of the Danks clusters is found, specifically $E(J-H)/E(H-Ks)=2.20\pm0.06$, which exceeds published values for the inner Galaxy. The G305 star forming complex, which includes the Danks clusters, lies beyond the Sagittarius-Carina spiral arm and occupies the Centaurus arm. Finally, the first deep infrared colour-magnitude diagram of RCW\,79 is presented, which reveals a sizeable pre-main sequence population. A list of candidate variable stars in G305 region is reported.}}
{\new{This study demonstrates the strength of the dataset and methodology employed, and constitutes the first step of a broader study which shall include reliable parameters for a sizeable number of poorly characterised and/or newly discovered clusters.}}


   \maketitle
%

\section{Introduction}

\begin{figure}[htbp]
  \centering
  \includegraphics[width=7.5cm]{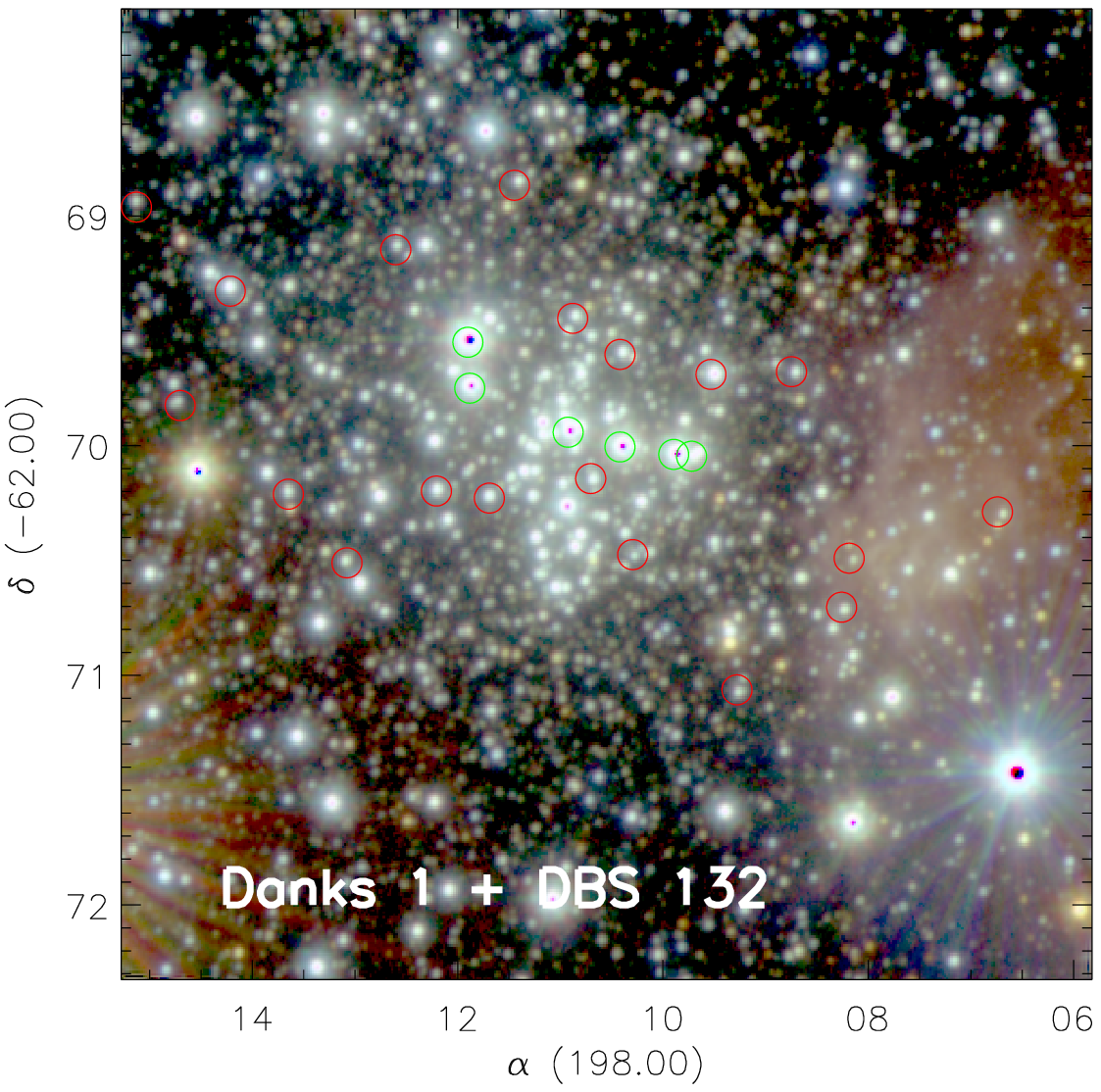}
  \includegraphics[width=7.5cm]{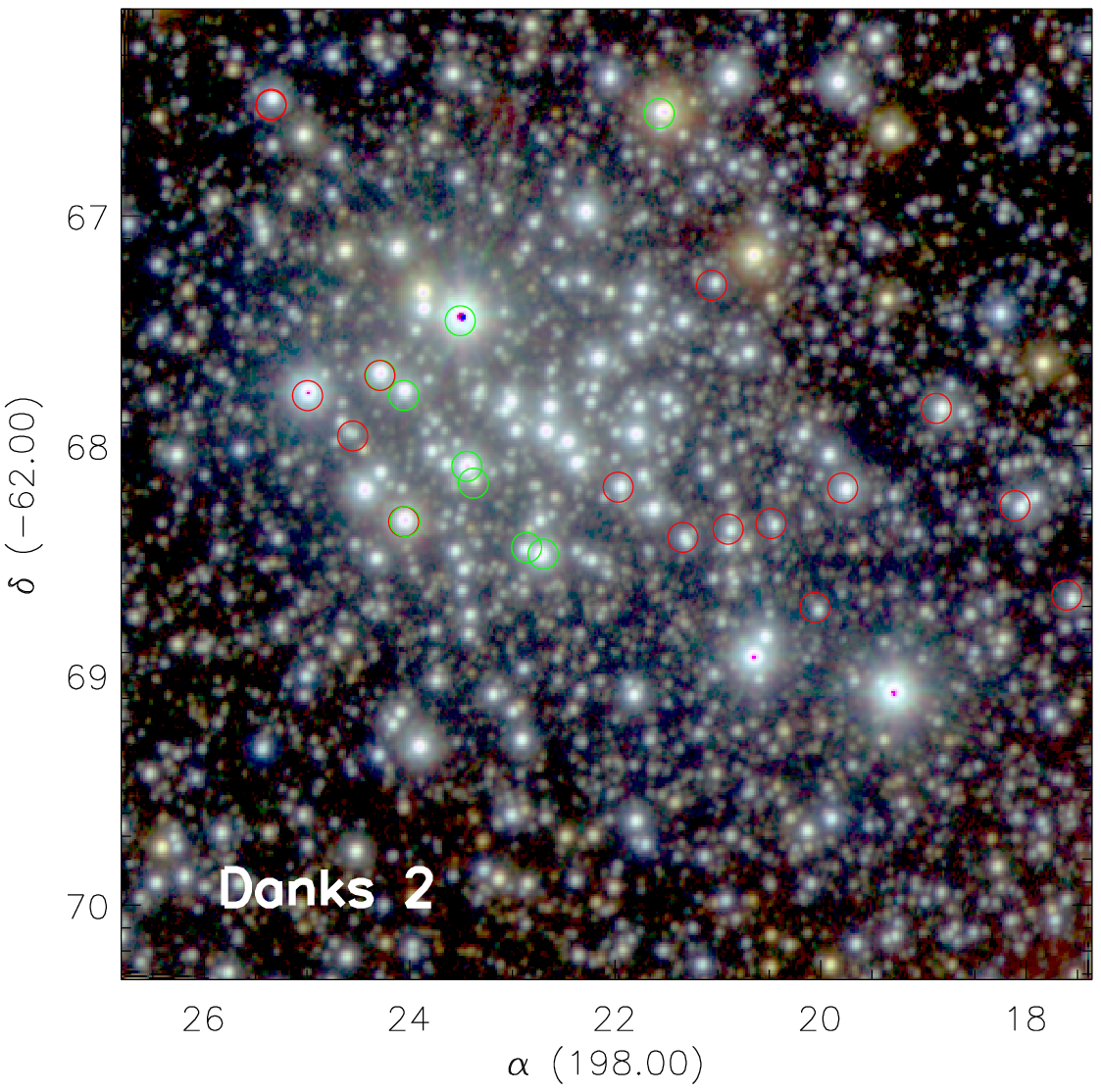}
  \includegraphics[width=7.5cm]{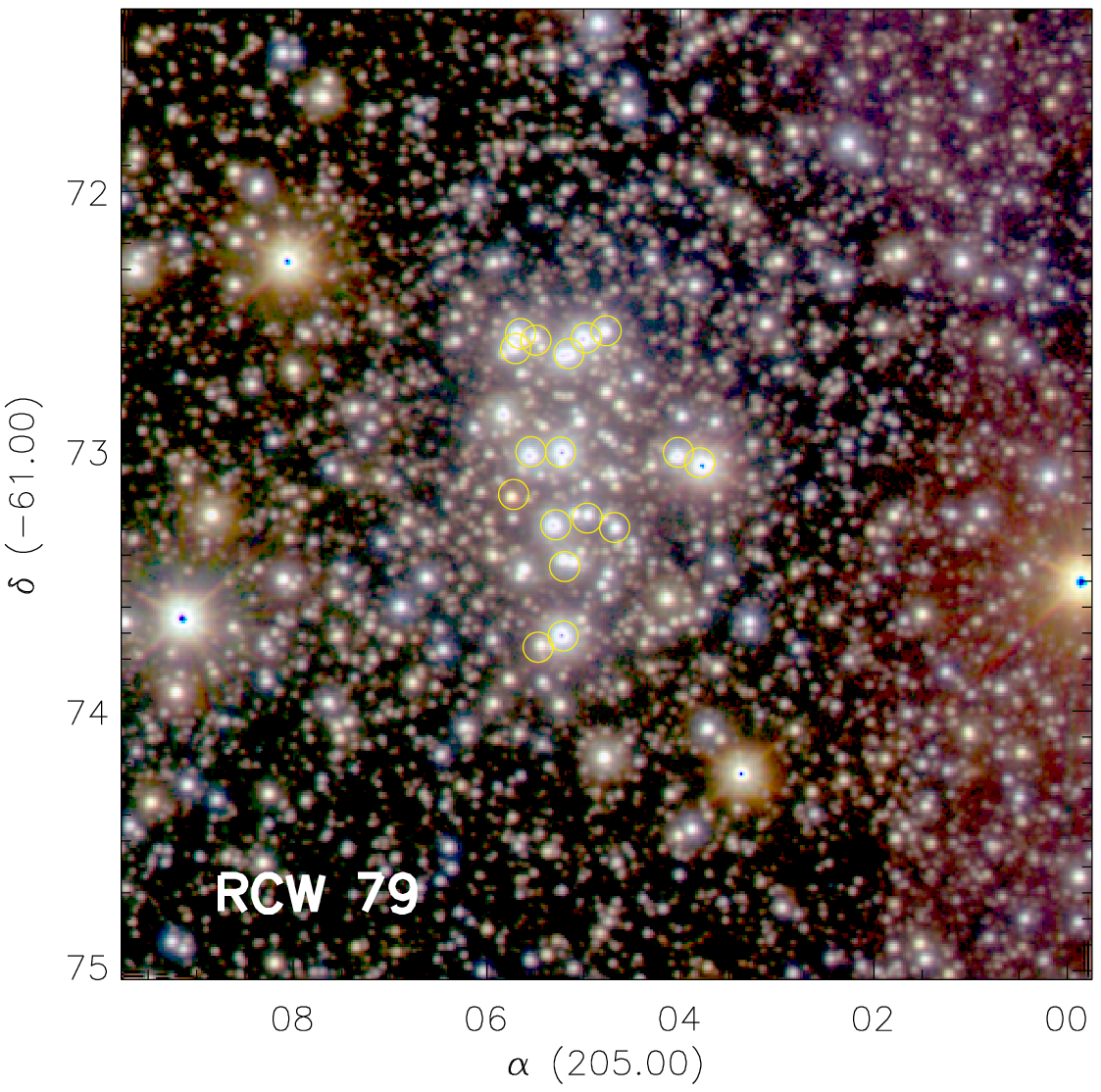}
      \caption{$JHK_s$ true colour images of the three clusters. Stars marked by circles were observed using near infrared spectrographs. In green are the stars observed using SofI at NTT and in red are the stars observed with MMIRS at the Clay telescope. The stars marked with a yellow circle in RCW\,79 were already observed by Martins et al. (\cite{Ma10}) and, hence, were not re-observed by us.}
        \label{FigTrue}
\end{figure}

It is commonly accepted that the majority of stars with masses  $\geq$ 0.50 M${_\odot}$  form in clustered environments (e.g. Lada \& Lada \cite{La03}, de Wit at al. \cite{Wi05}), rather than individually. Our location within our own Galaxy gives us a unique perspective from which we can study star clusters in great detail and such studies have important implications for our understanding of the formation of large galaxies in general. 

Estimates indicate that the Milky Way (MW) presently hosts  23000--37000 or more star clusters (Portegies Zwart et al. \cite{Po10}). However, only  2135 open clusters have been identified (according to the 26 Jan 2012 version of Dias et al. \cite{Di02}), which constitute a sample affected by several well known selection effects (as with globular clusters; Ivanov et al. \cite{Iv05}). Less than half of these clusters have actually been studied, and this subset suffers from further selection biases. Extending this current sample towards the Milky Way's highly obscured central region would provide unique insight into the formation, evolution, and dissipation of open clusters in the Galactic environment. To achieve this goal, we are using the unprecedented deep infrared data from the VISTA Variables in the V\'ia L\'actea (VVV) survey (Minniti et al. \cite{Mi10}, Saito et al. \cite{Sa12}), one of the six ESO Public Surveys operating on the new 4-meter Visible and Infrared Survey Telescope for Astronomy (VISTA). We are in the process of building a large sample of star clusters (including many discovered by our group; Borissova et al. \cite{Bo11}, Bonatto et al. in prep), that are practically invisible in the optical bands. The strength of this sample will lie in the homogeneity of the data (i.e. all observed with the same instrument and set-up) and analysis employed. From which, we will estimate clusters' physical parameters, including: angular sizes, radial velocities (RVs), reddening, distances, masses, and ages. Moreover, as pointed out by Majaess et al. (\cite{Maj12}), VVV photometry allows these parameters to be determined with unprecedented accuracy for highly obscured clusters. 

\new{As a first step}, we are focusing our efforts on young open clusters in their first few Myrs. During this period, which corresponds to Phase $I$ in the recent classification of Portegies Zwart et al. \cite{Po10}, stars are still forming and the cluster contains a significant amount of gas. The evolution of the cluster during this phase is governed by a complex mixture of gas dynamics, stellar dynamics, stellar evolution, and radiative transfer, and is currently not completely understood (Elmegreen \cite{El07}, Price \& Bate \cite{Pr09}). Thus many basic (and critical) cluster properties, such as the duration and efficiency of the star-formation process, the cluster survival probability and the stellar mass function at the beginning of the next phase are uncertain.

In this paper, we present a first sub-sample of 3 known young open clusters, studying them with both VVV colour-magnitude diagrams (CMDs) and low resolution near-infrared spectroscopy of the brightest stellar members. These clusters are RCW\,79, already studied by Martins et al. (\cite{Ma10}), and Danks\,1 and Danks\,2, discussed by Davies et al. (\cite{Da11}). We aim to describe our approach and present a first glimpse of the data quality. We detail our approach for determining the physical parameters of clusters observed with VVV, using previous works as references. As DBS\,132 is located close to Danks\,1 and Danks\,2, we also examine its CMDs and derive some preliminary cluster parameters. 

We begin by presenting the data in Section\,\ref{obs}, and relating our method and evaluating the accuracy of our work in Section\,\ref{res}. Subsequently, we describe the stellar variability detected in the clusters and their surroundings in Section\,\ref{var}. Then, in Section\,\ref{Discussion} we compare our results with previous studies and briefly discuss the characteristics of the star-forming regions in which the clusters are situated. Before concluding by summarising this work in Section\,\ref{Summary}.

\section{Observations}\label{obs}

\subsection{Photometry}\label{obs_phot}

\subsubsection{VVV data and photometry extraction}

We downloaded, from the VISTA Science Archive (VSA) website\footnote{http://horus.roe.ac.uk/vsa/}, the stacked images of the individual 2048$\times$2048 pixel exposures containing the three clusters presented in this paper. Danks\,1 and Danks\,2 both appear in VVV field d084, whilst RCW\,79 falls in VVV field d086. These fields were observed twice in the $ZYJHK_s$ bands on 7 to 29 March 2010 (d084), and 18 March to 4 April 2010 (d086), with the VIRCAM camera mounted on the VISTA 4m telescope at Paranal Observatory (Emerson \& Sutherland \cite{Em10}). The images were then reduced at CASU\footnote{http://casu.ast.cam.ac.uk/} by the VIRCAM pipeline v1.1 (Irwin et al. \cite{Ir04}). The total exposure time of each of these images was 40s, with 2 images per filter, on average. For a detailed description of the observing strategy see Minniti et al. (\cite{Mi10}); Saito et al. \cite{Sa12} provide further details about VVV data. 

During the observations the weather conditions fell within all the survey constraints for seeing, airmass, and Moon distance (Minniti et al. \cite{Mi10}) and the quality of the data was satisfactory. $JHK_s$ true colour images of the three clusters are shown in Figure~\ref{FigTrue}. Additional 8s $K_s$-band images were also obtained  in order to find and monitor variable stars in these fields. The dates, airmass, seeing, ellipticity and observation quality grade are all listed in Table~\ref{table:1} (see Online Material for complete version).

\begin{table*}
\caption{Observation log. Columns include UT date, HJD-2.455e6, airmass, seeing in arcsec, ellipticity and quality grade provided by ESO. Both the d084 and d086 fields are presented. (see Online Material for complete version)}             
\label{table:1}      
\centering                          
\begin{tabular}{rccccccrccccc}        
\hline\hline
\noalign{\smallskip}
\multicolumn{6}{c}{VVV field d084} && \multicolumn{6}{c}{VVV field d086}\\
\multicolumn{1}{c}{UT Date} & HJD-2.455e6 &  AM   & Sng & Ell. & QG && \multicolumn{1}{c}{UT Date} & HJD-2.455e6 &  AM   & Sng & Ell. & QG \\
\noalign{\smallskip}
\hline
\noalign{\smallskip}

29 Mar 2010 & 284.092893 & 1.550 & 0.91'' & 0.05 & A &&  23 Apr 2010 & 309.040379 & 1.552 & 0.67'' & 0.08 & A\\
29 Mar 2010 & 284.093200 & 1.547 & 0.85'' & 0.06 & A &&  23 Apr 2010 & 309.040738 & 1.548 & 0.69'' & 0.09 & A\\	
   7 Apr 2010 & 293.088925 & 1.462 & 0.89'' & 0.08 & C && 11 May 2010 & 327.158628 & 1.269 & 0.76'' & 0.17 & C\\
\multicolumn{1}{c}{...} & ... & ... & ... & ... & ... && \multicolumn{1}{c}{...} & ... & ... & ... & ... & ...\\
\noalign{\smallskip}
\hline
\end{tabular}
\end{table*}

\onltab{1}{
\begin{table*}
\caption{Observation log. Columns include UT date, HJD-2.455e6, airmass, seeing in arcsec, ellipticity and quality grade provided by ESO. Both the d084 and d086 fields are presented.}             
\label{table:1el}      
\centering                          
\begin{tabular}{rccccccrccccc}        
\hline\hline
\noalign{\smallskip}
\multicolumn{6}{c}{VVV field d084} && \multicolumn{6}{c}{VVV field d086}\\
\multicolumn{1}{c}{UT Date} & HJD-2.455e6 &  AM   & Sng & Ell. & QG && \multicolumn{1}{c}{UT Date} & HJD-2.455e6 &  AM   & Sng & Ell. & QG \\
\noalign{\smallskip}
\hline
\noalign{\smallskip}

29 Mar 2010 & 284.092893 & 1.550 & 0.91'' & 0.05 & A &&  23 Apr 2010 & 309.040379 & 1.552 & 0.67'' & 0.08 & A\\
29 Mar 2010 & 284.093200 & 1.547 & 0.85'' & 0.06 & A &&  23 Apr 2010 & 309.040738 & 1.548 & 0.69'' & 0.09 & A\\	
   7 Apr 2010 & 293.088925 & 1.462 & 0.89'' & 0.08 & C && 11 May 2010 & 327.158628 & 1.269 & 0.76'' & 0.17 & C\\
   7 Apr 2010 & 293.089231 & 1.459 & 0.94'' & 0.06 & C && 11 May 2010 & 327.158984 & 1.268 & 0.80'' & 0.15 & C\\
   7 Apr 2010 & 293.091056 & 1.454 & 1.12'' & 0.07 & C && 12 May 2010 & 328.150803 & 1.265 & 0.77'' & 0.13 & C\\
   7 Apr 2010 & 293.091364 & 1.451 & 0.92'' & 0.10 & C && 12 May 2010 & 328.151117 & 1.264 & 0.78'' & 0.11 & C\\
   7 Apr 2010 & 293.093232 & 1.446 & 0.95'' & 0.12 & C &&    9 Jun 2010 & 356.088306 & 1.277 & 0.92'' & 0.19 & C\\
   7 Apr 2010 & 293.093539 & 1.444 & 0.89'' & 0.10 & C &&    9 Jun 2010 & 356.088670 & 1.276 & 0.95'' & 0.25 & C\\
11 May 2010 & 327.149710 & 1.283 & 0.67'' & 0.12 & C && 16 Jun 2010 & 363.096838 & 1.317 & 0.76'' & 0.20 & C\\
11 May 2010 & 327.150016 & 1.282 & 0.73'' & 0.14 & C && 16 Jun 2010 & 363.097294 & 1.317 & 0.75'' & 0.15 & C\\
12 May 2010 & 328.137896 & 1.275 & 0.84'' & 0.06 & C && 21 Jun 2010 & 368.087491 & 1.326 & 0.93'' & 0.08 & A\\
12 May 2010 & 328.138247 & 1.274 & 1.13'' & 0.17 & C && 21 Jun 2010 & 368.087910 & 1.325 & 1.06'' & 0.03 & A\\
   8 Jun 2010 & 355.964570 & 1.328 & 0.91'' & 0.08 & C &&   13 Jul 2010 & 390.036517 & 1.346 & 0.79'' & 0.09 & C\\
   8 Jun 2010 & 355.964937 & 1.326 & 0.93'' & 0.14 & C &&   13 Jul 2010 & 390.036977 & 1.346 & 0.80'' & 0.05 & C\\
 16 Jun 2010 & 363.080558 & 1.326 & 0.74'' & 0.16 & C && 15 Aug 2010 & 423.990697 & 1.499 & 0.72'' & 0.11 & B\\
 16 Jun 2010 & 363.080968 & 1.325 & 0.73'' & 0.22 & C && 15 Aug 2010 & 423.991116 & 1.500 & 0.75'' & 0.08 & B\\
 21 Jun 2010 & 368.063868 & 1.320 & 0.92'' & 0.11 & C && 17 Aug 2010 & 425.010078 & 1.607 & 0.69'' & 0.07 & A\\
 21 Jun 2010 & 368.064288 & 1.319 & 0.95'' & 0.12 & C && 17 Aug 2010 & 425.010549 & 1.609 & 0.74'' & 0.08 & A\\
    8 Jul 2010 & 385.063093 & 1.439 & 0.76'' & 0.07 & B &&  15 May 2011 & 696.219013 & 1.402 & 0.60'' & 0.13 & A\\
    8 Jul 2010 & 385.063579 & 1.439 & 0.77'' & 0.05 & B &&  15 May 2011 & 696.219447 & 1.402 & 0.57'' & 0.11 & A\\
    9 Jul 2010 & 386.054580 & 1.419 & 0.62'' & 0.13 & A &&    16 Jul 2011 & 758.067592 & 1.464 & 1.15'' & 0.11 & C\\
    9 Jul 2010 & 386.055029 & 1.420 & 0.57'' & 0.08 & A &&    16 Jul 2011 & 758.068015 & 1.465 & 1.22'' & 0.09 & C\\
  10 Jul 2010 & 387.063990 & 1.462 & 0.60'' & 0.12 & A &&    30 Jul 2011 & 772.017268 & 1.420 & 0.69'' & 0.05 & A\\
  10 Jul 2010 & 387.064418 & 1.462 & 0.64'' & 0.11 & A &&    30 Jul 2011 & 772.017713 & 1.421 & 0.73'' & 0.13 & A\\
10 Aug 2010 & 418.007474 & 1.588 & 0.79'' & 0.19 & A &&    31 Jul 2011 & 773.009639 & 1.404 & 1.37'' & 0.07 & N/A\\
10 Aug 2010 & 418.007953 & 1.589 & 0.79'' & 0.14 & A &&    31 Jul 2011 & 773.010068 & 1.405 & 1.35'' & 0.08 & N/A\\
 14 Jun 2011 & 726.145651 & 1.502 & 0.73'' & 0.07 & A &&    2 Aug 2011 & 775.013248 & 1.435 & 2.30'' & 0.19 & C\\
 14 Jun 2011 & 726.146109 & 1.502 & 0.78'' & 0.07 & A &&    2 Aug 2011 & 775.013651 & 1.435 & 2.02'' & 0.20 & C\\
  24 Jul 2011 & 766.025934 & 1.460 & 0.74'' & 0.10 & A &&    4 Aug 2011 & 777.022491 & 1.491 & 0.77'' & 0.11 & B\\
  24 Jul 2011 & 766.026379 & 1.460 & 0.75'' & 0.06 & A &&    4 Aug 2011 & 777.022898 & 1.492 & 0.70'' & 0.10 & B\\
  7 Aug 2011 & 780.991745 & 1.486 & 0.80'' & 0.13 & A &&  15 Aug 2011 & 788.017516 & 1.614 & 0.85'' & 0.09 & B\\
  7 Aug 2011 & 780.992151 & 1.487 & 0.88'' & 0.09 & A &&  15 Aug 2011 & 788.017933 & 1.616 & 0.76'' & 0.07 & B\\
16 Aug 2011 & 789.979686 & 1.541 & 0.88'' & 0.14 & A && 17 Aug 2011 & 790.993083 & 1.531 & 1.16'' & 0.07& N/A\\
16 Aug 2011 & 789.980120 & 1.542 & 0.96'' & 0.04 & A && 17 Aug 2011 & 790.993548 & 1.532 & 1.19'' & 0.05 & N/A\\
                        &                        &            &            &          &     && 21 Aug 2011 & 794.993380 & 1.587 & 0.88'' & 0.11 & A\\
                        &                        &            &            &          &     && 21 Aug 2011 & 794.993802 & 1.588 & 0.90'' & 0.05 & A\\
                        &                        &            &            &          &     && 22 Aug 2011 & 795.995700 & 1.615 & 1.56'' & 0.05 & C\\
                        &                        &            &            &          &     && 22 Aug 2011 & 795.996109 & 1.616 & 1.61'' & 0.09 & C\\
\noalign{\smallskip}
\hline
\end{tabular}
\end{table*}
}

\begin{figure}
  \centering
   \includegraphics[width=9.0cm]{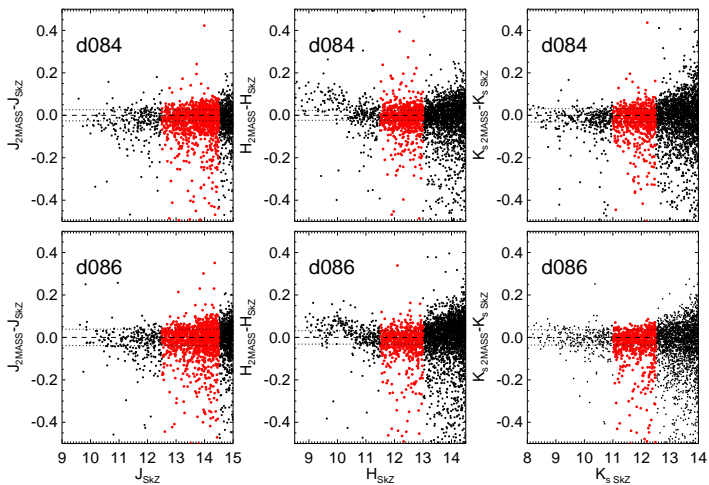}
   \caption{Comparison between the VVV-SkZ\_pipeline's $JHK_s$ photometry and that from the 2MASS catalogue. Since the VVV-SkZ\_pipeline's astrometric solution is based on the 2MASS catalogue, we were able to match most of the stars within a 1''-radius. Red, big circles are the stars within the magnitude intervals used for the flux calibration, i.e. $12.5~<J<14.5$~mag, $11.5<H<13$~mag and $11<K_s<12.5$~mag. A dashed line shows where the VVV-SkZ\_pipeline's photometry perfectly matches the 2MASS values, and the dotted lines show the standard deviation of the points.}
              \label{FigCal}
\end{figure}

The VVV-SkZ\_pipeline (Mauro et al. 2012) was employed to determine stellar photometry from the images. The VVV-SkZ\_pipeline is automated, based on ALLFRAME (Stetson
\cite{St94}) and optimised for PSF photometry of VVV data. The VVV-SkZ\_pipeline uses single pointing images, called paw prints
(covering 0.59 sq. deg). CASU aperture photometry is available for paw prints as well as tiles, which are the combination of 6 pawprints giving a full 1.64 sq. deg combined image (see Saito et al. \cite{Sa12} for more details about pawprints and tiles). For PSF photometry we only use paw prints as sources on a tile have PSF shapes that are too difficult to model, due to the variable PSF across each combined tile. In such conditions PSF fitting does is no more accurate than simple aperture photometry. To take into account the variable PSF, the VVV-SkZ\_pipeline simply allows the PSF to vary quadratically across the pawprintsÕ field-of-view.

The final product of the pipeline is an astrometric, flux calibrated star catalogue. \new{PSF photometry produces many spurious detections close to defects on the detectors, in the border of the fields, in the core and on the speckles of saturated stars. These can be isolated when, for all sources, the photometric errors given in ALLFRAME's output are compared with the apparent magnitudes, since the error on the fit of fake detections tends to be comparatively high. The VVV-SkZ\_pipeline finds the maximum of the error distribution per magnitude bin (typically 0.1~mag), and fits a polynomial function to get the photometric error as a function of magnitude. We then employ a sigma clipping algorithm to remove spurious detections: all detections showing a photometric error higher than 3$\sigma$ are removed. One should note that the errors given by ALLFRAME are not the real photometric error, since it does not take into account all sources of error. For instance, it does not account for the correlation in the noise between adjacent pixels in infrared detectors.}

Calibration  in the VVV $J$-, $H$- and $K_s$-bands was done using 2MASS stars with $12.5<J<14.5$~mag, $11.5<H<13$~mag and $11<K_s<12.5$~mag (see Figure~\ref{FigCal}). Stars brighter than $K_s \sim 12.5$~mag are saturated in the VVV disk area, however ALLFRAME uses the wings of the point-spread function of these stars to estimate their apparent magnitude, and we obtain a relative accuracy of 0.05 mag for stars up to mag=9.0.  For stars brighter than $K_s$=9.0~mag we just adopted the 2MASS magnitudes. At the faint end a relative accuracy of 0.1~mag or better is achieved down to $J$=20.0~mag or $K_s$=18.0~mag. We did not correct the magnitudes of the 2MASS calibration stars into the VISTA photometric system, or apply any extinctions related terms (see below) unlike the procedure adopted by the VISTA Data Flow System pipeline at CASU. Thus our magnitudes differ from those in the catalogues produced by CASU which are available through the VSA and via ESO\footnote{http://archive.eso.org/wdb/wdb/adp/phase3\underline{ }vircam/form}. We designate our VVV magnitudes in the 2MASS system as M(VVV$_{2MASS}$) to distinguish them from magnitudes for the same objects in the VISTA system, M(VISTA). We opted to work in the 2MASS system rather than remaining in the natural VISTA system, as most of the models we are using in this study, and most of the other studies of massive clusters, employ the 2MASS photometric system. 

For each of the 16 detectors in each pawprint our calibration of VVV data uses the 2MASS magnitude M(2MASS)  of stars with colours col(2MASS)  and the instrumental magnitudes m(VVV) from the VVV-SkZ\_pipeline psf fitting pipeline. It determines the two constants M$_0$ the zero-point and $d$ the colour coefficient in the fit M(2MASS) $-$ m(VVV) $=$ M$_0+d$\,col(2MASS). The magnitudes of the VVV sources on the 2MASS system M(VVV$_{2MASS}$) are then found by inverting this equation. Our full transformation equations also include aperture correction and airmass effects. 

Of course, transformations between two photometric systems depend on reddening. As we calibrate flux separately for each VIRCAM detector, we are effectively assuming, through our colour coefficient $d$, a mean reddening in each of those chips. One could use Schlegel et al. (\cite{Sc98})'s reddening maps, but this map's spatial resolution is equivalent to nearly one VIRCAM chip and should provide similar results. Moreover, the use of these maps is also highly questionable at these latitudes as mentioned by Schlegel et al. (\cite{Sc98}) themselves, since at low Galactic latitudes ($|b|<5^\circ$), most contaminating sources have not been removed from their maps, and the temperature structure of the Galaxy is not well resolved. Furthermore, no comparisons between our predicted reddenings from $d$ and observed reddening have been made in these regions. One should note that the 2MASS photometry shows a dispersion higher than 0.1~mag for $K>13$~mag. 

To calculate the completeness of our catalogue, we created Luminosity Functions (LFs) for the clusters, with a bin size of 0.5 $K$-band magnitudes. For each magnitude bin, we added artificial stars within the given magnitude ranges to the cluster image. The completeness was calculated by finding the recovery fraction of artificial stars per magnitude. The detection of the stars was done using {\sc dao}phot following a similar methodology as performed by the VVV-SkZ\_pipeline. 

\subsubsection{Field star decontamination}\label{decont}

\new{To disentangle field and cluster stars we employed the statistical} decontamination algorithm described in Bonatto \& Bica (\cite{Bon10}, and references therein), adapted to exploit the greater photometric depth of VVV. The algorithm first requires the identification of a comparison field. Depending on the projected distribution of individual stars, clusters or clumpy extinction, this may be a ring around the cluster or some other region selected in its immediate vicinity. The algorithm divides the full range of CMD magnitudes \new{and colours into a grid of cells with axes along $K_s$, ($H - K_s$) and ($J-K_s$). Initial cell dimensions are $\Delta K_s=1.0$ and $\Delta$($H - K_s$) = $\Delta$($J - K_s$) = 0.2~mag. However, sizes half and twice those values are also considered, together with shifts in the grid positioning by $\pm$1/3 of the respective cell size along the 3 axes. Thus, 729 independent decontamination outputs were obtained for each cluster candidate. For each cell, the algorithm estimates the expected number density of member stars by subtracting the respective field star number-density\footnote{Photometric uncertainties are considered by computing the probability of a star of given magnitude and colour to be found in any cell (i.e., the difference of the error function computed at the cell borders).}. Summing over all cells, each grid setup produces a total number of member stars $N_{\rm{mem}}$ and, repeating this procedure for the 729 different setups, we obtain the average number of member stars $\langle N_{\rm{mem}}\rangle$. Each star is ranked according to the survival frequency after all runs, and only the $\langle N_{\rm{mem}}\rangle$ highest ranked stars are taken as cluster members. A full description of the algorithm is given in Bonatto \& Bica (\cite{Bon10}). As a photometric quality constraint, the algorithm rejects stars with $K_s$ or colour uncertainties greater than 0.1~mag.}

\begin{figure}
  \centering
   \includegraphics[width=7.5cm]{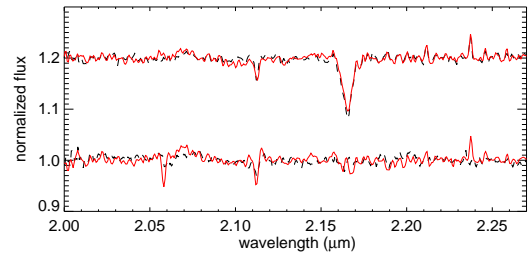}
   \caption{Section of the $K_s$ spectrum of two stars that were observed using different instruments and/or different slit masks. Obj2 (top) was observed with MMIRS using the first (solid, black line) and the second (dashed, red line) slit mask. Obj5 (bottom) was observed with NTT/SofI (solid, black line) and Clay/MMIRS (red, dashed line). An arbitrary shift in flux was applied for better visual representation.}
              \label{FigSpCmp}%
\end{figure}

\begin{figure*}[ht]
  \centering
   \includegraphics[width=6.0cm]{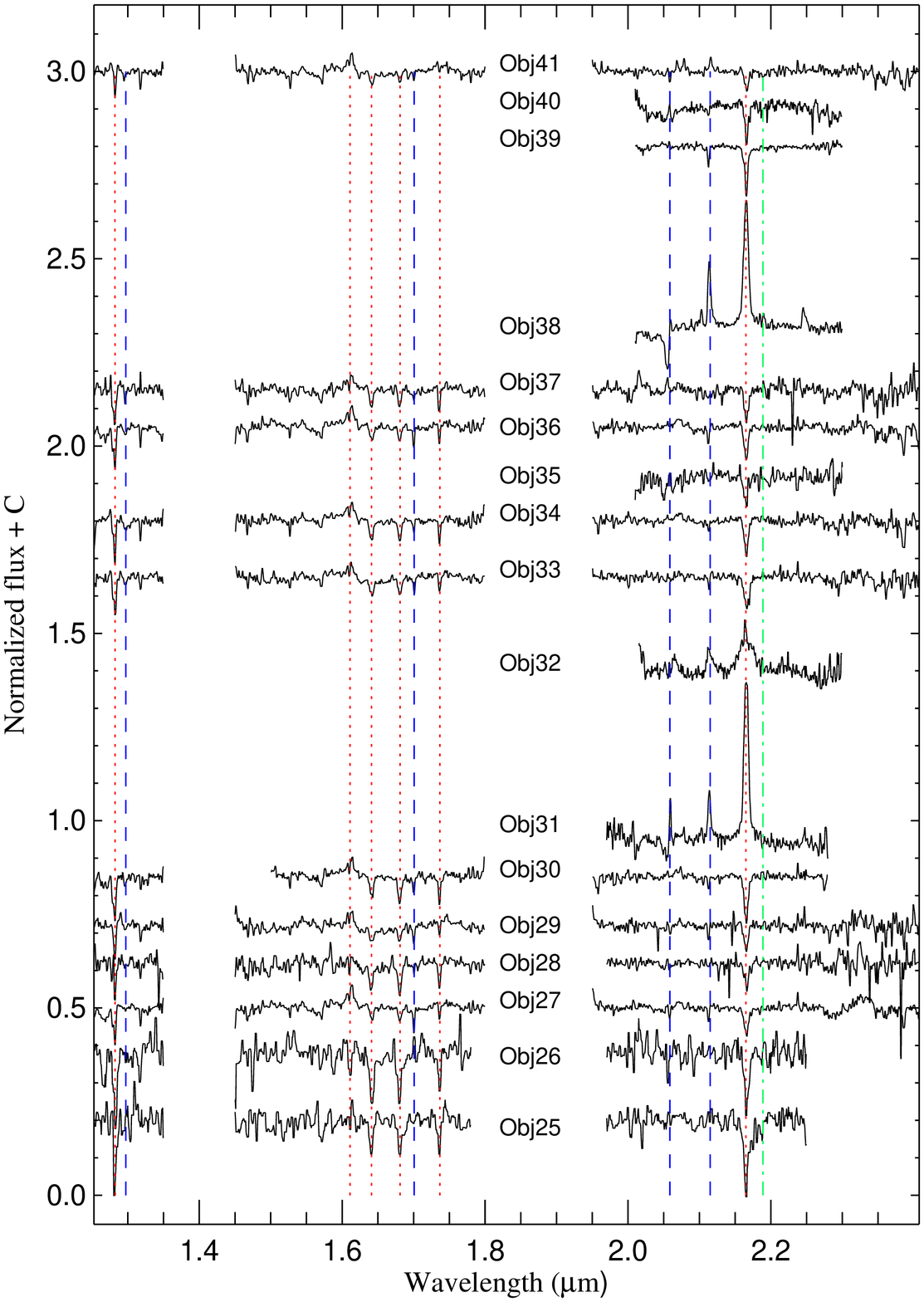}
   \includegraphics[width=6.0cm]{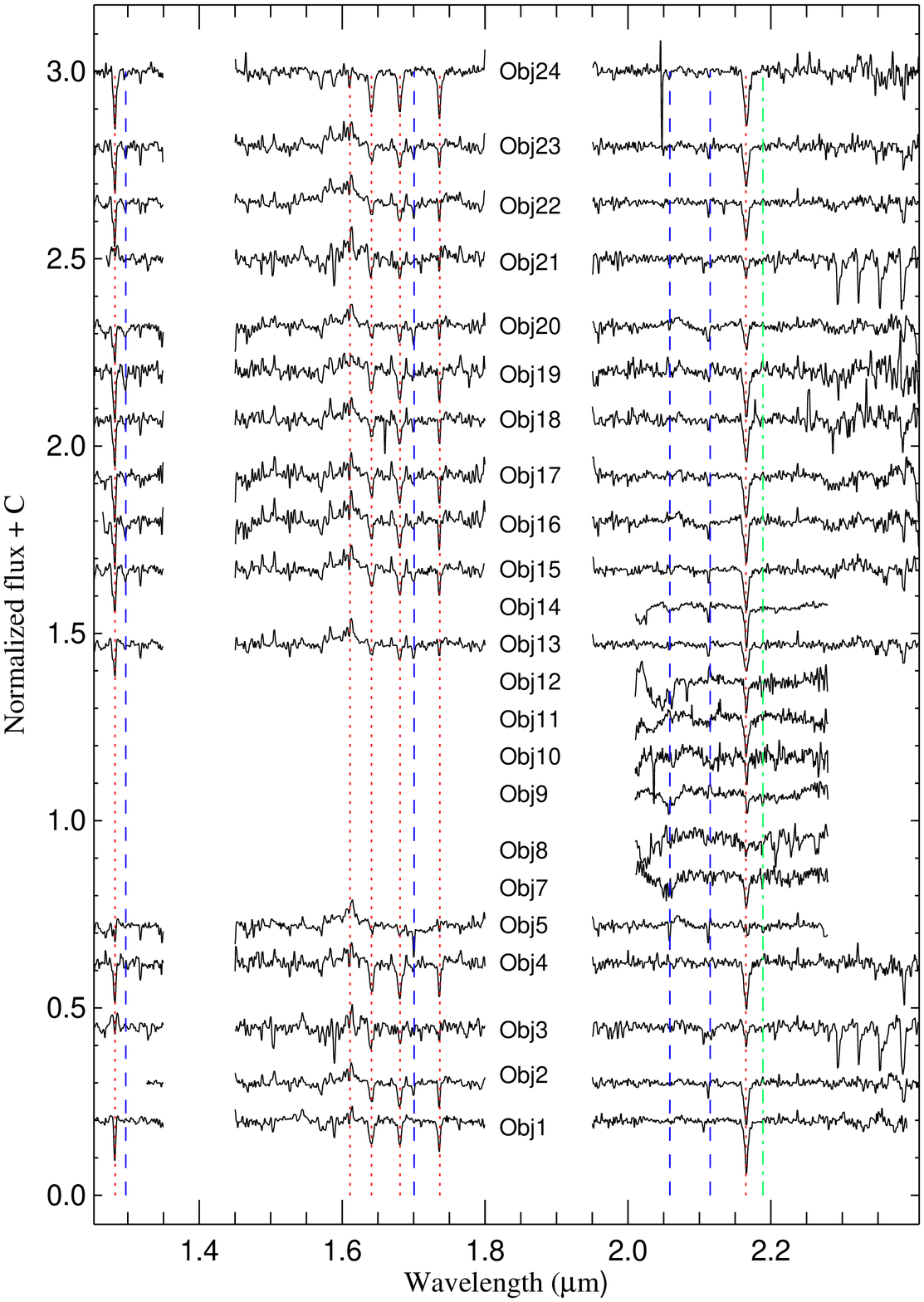}
   \includegraphics[width=6.0cm]{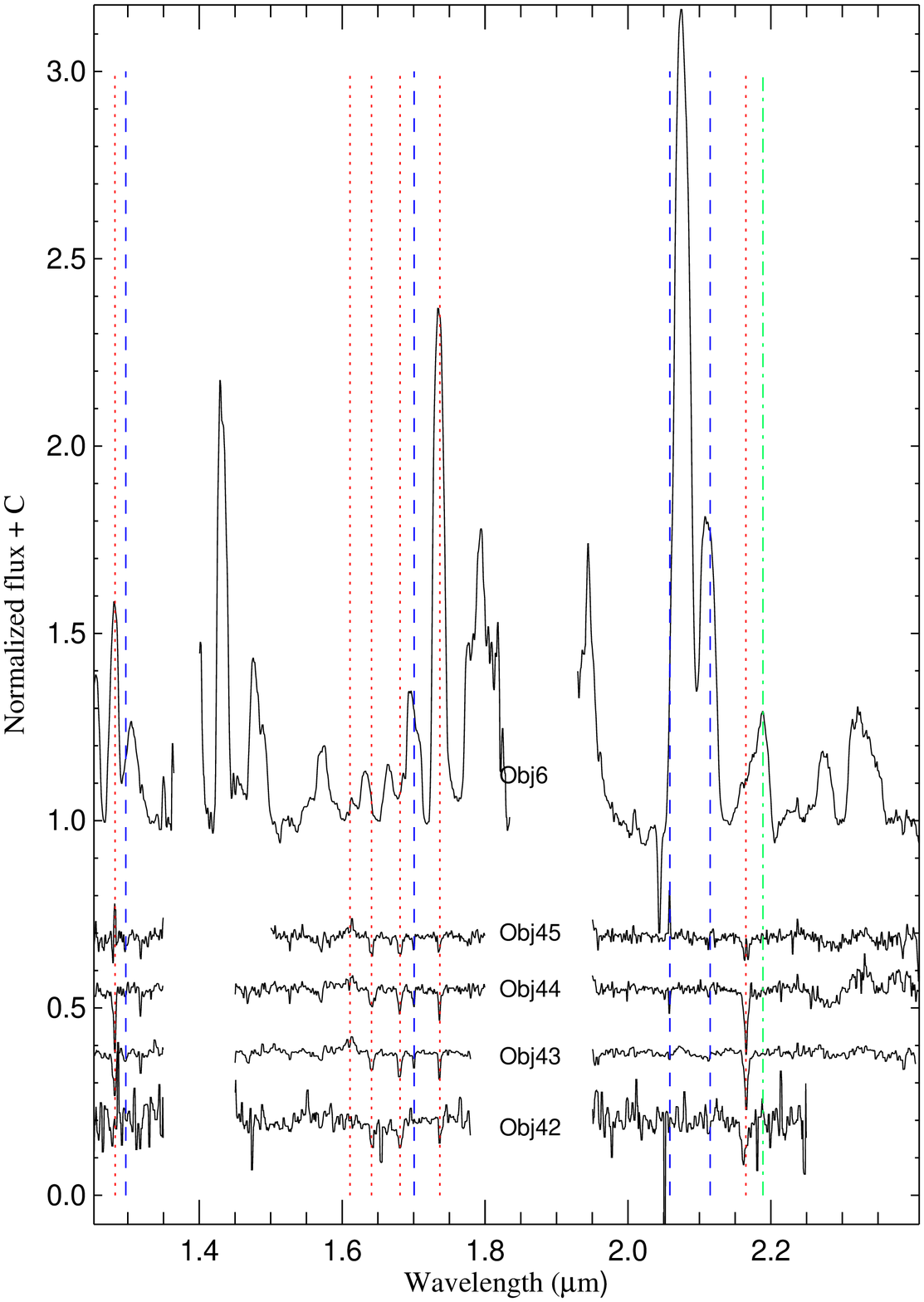}
   \caption{Spectra of the stars in the area of the clusters Danks\,1 (left), Danks\,2 (middle) and DBS\,132 (right). Obj6 (WC8 type) should be placed with the Danks\,2's spectra, but is presented in the right panel for better visibility. Hydrogen recombination lines are marked with vertical dotted, red lines and Helium lines with vertical dashed, blue lines.}
              \label{FigSp1}%
\end{figure*}

\subsection{Near-infrared spectroscopy}

\subsubsection{Observations}
We collected spectra of 6  members of Danks\,1 and 8 members of Danks\,2, using the infrared spectrograph and imaging camera SofI in long-slit mode on the New Technology Telescope (NTT) at La Silla Observatory ESO, Chile. These stars are marked in Figure~\ref{FigTrue} by green circles. Using the medium resolution grism in the 3$\rm^{rd}$ order, we covered the whole $K_s$-band, 2.00-2.30 $\mu$m, with a resolution of $\Delta\lambda$=4.66\,\AA/pix. We used a 1 arcsec slit, in order to match the seeing, which gives a resolving power of R$\sim$1320. For optimal subtraction of the atmospheric OH emission lines, we used 60 arcsec nodding along the slit in an ABBA pattern: the star was observed before (A) and after (B) a first nod along the slit, then at position B a second time before returning to position A for a last exposure. All stars were observed using two slit positions per cluster. The position angle of the slit was chosen so that two or more stars were observed simultaneously. Total exposure time was 40s for stars in Danks\,1, and 160s and 600s for the brightest and the faintest stars in Danks\,2, respectively. The total signal-to-noise ratio (S/N) per pixel ranges from 60 for the faintest stars to 150 for the brightest. Bright stars of spectral type G were observed as a measure of the atmospheric absorption. These stars were selected to share the same airmass as the targeted clusters' stars during the middle of their observation. The spectra of the two stars HD\,113376 (G3\,V) and HD\,120954 (G1\,V) were obtained after the observations of the first and the second slit position on Danks\,1 on 14 and 18 April 2011, respectively. Also, the star HD\,119550 (G2\,V) was observed just after the two slit positions on Danks\,2 on 15 April 2011. 

We also obtained a total of 38 spectra of stars within and in the vicinity of the star clusters Danks\,1 and Danks\,2, using the near-infrared spectrograph MMIRS in multi-object mode on the Clay telescope at Las Campanas Observatory on 2011 May 19--20. These stars are indicated in Figure~\ref{FigTrue} by red circles. Using the HK grism, we covered a spectral range of $\lambda\lambda$ = 1.25--2.45 $\mu$m, with a resolution of $\Delta\lambda$=6.70\,\AA/pix. We needed two slit-mask designs to observe all our targets of interest. Both used a slit-width of 0.5 arcsec, which gives a resolving power of R$\sim$1120. As with the SofI data, nodding along the slits was used, but due to the small size of the slits, we only used 2 arcsec nodding in an ABCCBA pattern, which is similar to an ABBA pattern, but with third position along the slit. Individual exposure time was limited to 300s to allow for accurate subtraction of the sky emission and the total exposure time was 2700s for all stars. The average total S/N per pixel reached ranges from 100 to 200 for stars brighter than $K_s$ = 12.0~mag and from 30 to 90 for stars fainter than  $K_s$ = 12.0~mag. The star HD\,114012 (spectral type A0\,V) was observed after the first mask as a measure of the atmospheric absorption. 

Martins et al. (\cite{Ma10}) obtained high quality NIR spectra of stars in RCW79. Hence, we use their spectra to complement our photometry.

\subsubsection{Reduction and extraction}

Before standard procedures could be applied to reduce the data, some initial processing steps were required to deal with peculiarities of the instruments we used. In the case of SofI spectra, we first had to correct for bad pixels, using the latest bad pixel mask available on the European Southern Observatory webpage\footnote{http://www.eso.org/lasilla/instruments/sofi/tools/\\reduction/bad\_pix.html}. Then correct for cross-talk, as described in the SofI User's Manual (Moorwood et al. \cite{Mo98}). With MMIRS spectra, which were obtained through ``up the ramp'' acquisition, a few more steps were needed. Each fits file consists of a data cube containing the non-destructive readout of the whole detector at every 5s step. Hence, after subtracting the dark frame from the science data, one can collapse all readouts taken during an exposure into a single image by fitting a slope to the number of counts as a function of time for each pixel (as suggested on the MMIRS webpage\footnote{http://hopper.si.edu/wiki/mmti/MMTI/MMIRS/MMIRS+Pipeline}). By doing so, when the linearity and the saturation levels of each pixel are taken into account adequately, one corrects for cosmic rays and for saturated lines. The last step before running a standard reduction was to find, trace and cut the individual slits in the original image. The traces were fitted using a 4$^{th}$-order polynomial fit and rectified before being cut. All steps described above were executed using custom-written Interactive Data Language (IDL) scripts.

Subsequent nodding observations were subtracted from one another to remove bias level and sky emission lines. The flat fielding, spectrum extraction and wavelength calibration of all spectra were executed in the usual way using {\sc iraf}\footnote{{\sc iraf} is distributed by the National Optical Astronomy Observatories (NOAO), which is operated by the Association of Universities for Research in Astronomy, Inc. (AURA) under cooperative agreement with the U.S.A. National Science Foundation (NSF).}. Calibration lamp spectra (Xenon-Neon for SofI and Argon for MMIRS), taken at the beginning of each night, were used for wavelength calibration. The wavelength solution of each frame has an r.m.s. \new{uncertainty} of $\sim$0.5 pixels for SofI spectra and 0.2 pixels for MMIRS spectra, which correspond to $\sim$30 and $\sim$20~km~s$^{-1}$, respectively. Atmospheric absorption was corrected for with the {\sc iraf} task {\it telluric}. When a G-type star was used, it was first divided by a solar spectrum downloaded from the National Solar Observatory (Livingston \& Wallace \cite{Li91}) and resampled following the method described by Maiolino et al. (\cite{Ma96}). When an A0 star was used, we first subtracted a fitted Voigt profile from the standard Br absorption features before running the task {\it telluric}. 

Finally, all spectra were rectified using a low-order polynomial fit to a wavelength interval that was assumed to be pure continuum, i.e. without absorption or emission lines. Unfortunately, due to bad weather, observing was abruptly interrupted just before we could observe a telluric standard for the second mask at Las Campanas. Both masks were observed during different nights, but since they were observed at comparable airmasses ($\sim$1.2), we decided to use the same telluric standard for all spectra. Depending on how the weather conditions and the sky level were different during the two nights, this could have introduced uncertainties in the extraction of the spectra of the second mask, but nothing significant is observed. Indeed, one star was observed with both masks and, as one can see in Figure~\ref{FigSpCmp}, the two spectra are very similar. Figure~\ref{FigSpCmp} also shows the similarity of two spectra of another star observed using both SofI and MMIRs. Hence, our reduction and extraction methods can be trusted. 

\subsubsection{Spectral classification}\label{spcl}

The spectra of the stars in the area of the clusters Danks\,1, Danks\,2 and DBS2003\,132 are plotted in Figure\,\ref{FigSp1}. Preliminary spectral classifications were made using available catalogues of $K$-band spectra of objects with spectral types derived from optical studies (Morris et al. \cite{Mo96}; Figer et al. \cite{Fi97}; Hanson et al. \cite{Ha96,Ha05}) as well as the spectral catalogues of Martins \& Coelho (\cite{Ma07}), Crowther et al. (\cite{Cr06}), Liermann et al. (\cite{Li09}), Mauerhan et al. (\cite{Ma11}) and Davies et al. (\cite{Da11}). The most prominent lines in our wavelength range are: Br\,$\gamma$ (4--7) 2.1661 $\mu$m; Br\,10 (4--10) 1.737 $\mu$m; Br\,11 (4--11) 1.681 $\mu$m and  Br\,12 (4--12) 1.641 $\mu$m (all from the Brackett series); Pa\,$\beta$ (3--5) 1.282 $\mu$m, from the Paschen series; He{\sc i} lines at 2.1127 $\mu$m ($3p\,^3P^o$ -- $4s\,^3S$, triplet), 1.702 $\mu$m ($1s^4d^3D^e$ -- $1s^3p^3P^o$), 1.668 $\mu$m (4--11) and 1.722 $\mu$m (4--10); He{\sc ii} lines at 2.188 $\mu$m (7--10) and 1.692 $\mu$m (12--7); the blend of C{\sc iii}, N{\sc iii} and O{\sc iii} at 2.115 $\mu$m; N{\sc iii}\,$\lambda$2.103 $\mu$m; C{\sc iv} lines at 2.070 $\mu$m and 2.079 $\mu$m. All these lines were compared with the template spectra from the listed papers.

The Equivalent Widths (EWs) of the Br\,11 (4--11) and He{\sc i} 1.702 $\mu$m lines of the spectral targets were measured, on the continuum normalised spectra, by the {\sc iraf} task {\it splot}. \new{In general, we fit line profiles using a Gaussian with a linear background, in more complicated profiles we used the deblending function or Voigt profile fitting.} The EW uncertainties were estimated taking into account the S/N of the spectra, the peak to continuum ratio of the line (see Bik et al. \cite{Bi05}) and the error from the telluric star subtraction (which was estimated to be $\sim$10-15\% in the worst cases). The EWs of the emission lines are negative by definition. Their ratio was used for qualitative  estimation of the spectral types of OB stars, using the calibrations given in Hanson et al. (\cite{Ha98}). 

In general, the spectra of most stars show the He{\sc i} lines and the Brackett series in absorption, which is typical for O and early B main sequence (MS) stars. Several stars, namely Objs.\,1, 4, 19, 25, 26, 28 and 42, show only the Brackett series in absorption and are classified as later B - early A type stars. The stars Obj\,3 and 21 present CO, Mg{\sc i}, Fe, Ca and Al{\sc i} lines, characteristic of late type stars. Obj\,3 was observed by Davies at al. (\cite{Da11}, their number D2-2) and classified as F8/G1\new{, but they concluded that it is in the foreground from it's near-infrared colours}. We assign G7/8 I spectral type to Obj\,21 based on a comparison with the template spectral library of Rayner et al. (\cite{Ra09}). Objs. 6, 31, 32, 38 are WR stars previously discovered by Davies et al. (\cite{Da11}, their numbers D2-3, D1-1, D1-5 and D1-2) and Mauerhan et al. (\cite{Ma11}, their numbers MDM8 and MDM7). The adopted spectral types are given in Table~\ref{table:2}. This method of spectral classification is correct to within 2 subtypes in the near-IR. We adopted this error for our estimates.

\begin{figure}
  \centering
   \includegraphics[width=9.0cm,angle=0]{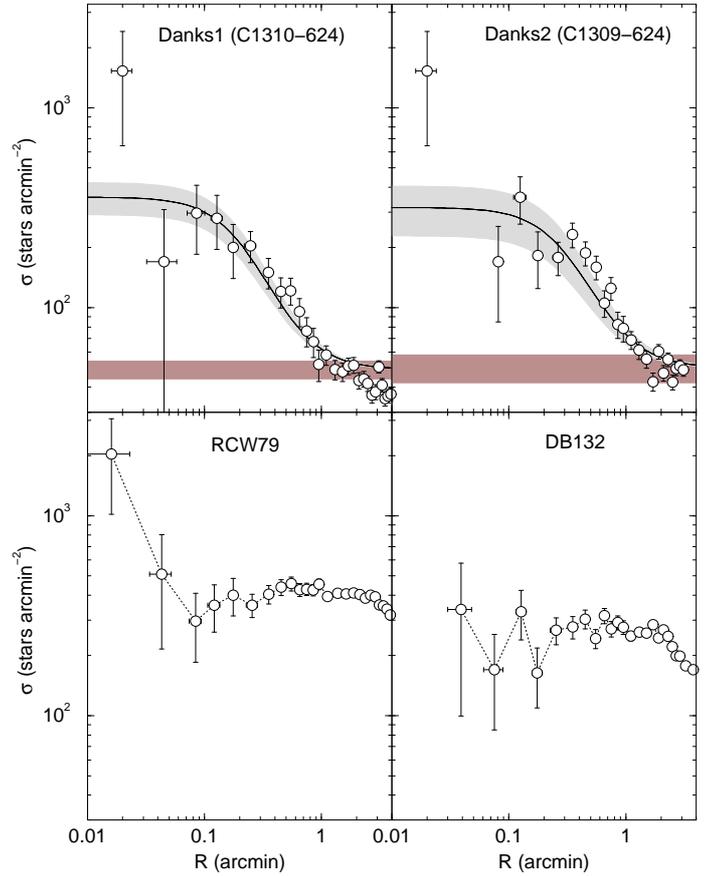}
   \caption{Radial density profiles for Danks\,1, Danks\,2 and RCW\,79. The error bars are given by Poissonian noise, the solid line shows the fit of a 2-parameter King profile (with uncertainties plotted in grey) and the purple rectangle gives the field density level. Unfortunately, no fit was possible for RCW\,79 and DBS\,132.}
              \label{FigRad}%
\end{figure}

\section{Results}\label{res}

In this section we describe our methods to determine the fundamental parameters of the young massive clusters, such as the angular size, RV, extinction, distance, age, and mass.

\subsection{Angular size}

One parameter that can be directly determined is the angular radius of the clusters. We first obtained the coordinates of the cluster centre, based on our stellar catalogue. Starting from a first-guess value, we calculated the radial density profile (RDP) iteratively until we \new{converged} towards a best profile. Using this method, we get:
\begin{eqnarray}
\nonumber
\rm{Danks\,1: RA (J2000)} = 13:12:26.74\\
\nonumber
\rm{DEC (J2000)} = -62:41:37.81\\
\nonumber
\rm{Danks\,2: RA (J2000)} = 13:12:55.14\\
\nonumber
\rm{DEC (J2000)} = -62:40:52.00\\
\nonumber
\rm{RCW\,79: RA (J2000)} = 13:40:11.27\\
\nonumber
\rm{DEC (J2000)} = -61:43:52.10
\end{eqnarray}

Once the converged RDP is obtained (see results in Figure\,\ref{FigRad}), we determine the stellar density of the field stars in the vicinity of the clusters. We fit a 2-parameter King profile to each cluster to guide our determination of their angular size. In contrast to King \cite{Ki66}, we fit the profile to star counts rather than surface brightness. The King profile is defined by 2 parameters, the projected central density of stars (S$_0$) and the core radius (r$_{\rm{C}}$). The uncertainties on the fit are given by the maximum and minimum values allowed within the error bars of the data points (stemming from Poissonian noise).

\begin{figure*}
  \centering
   \includegraphics[width=9.0cm]{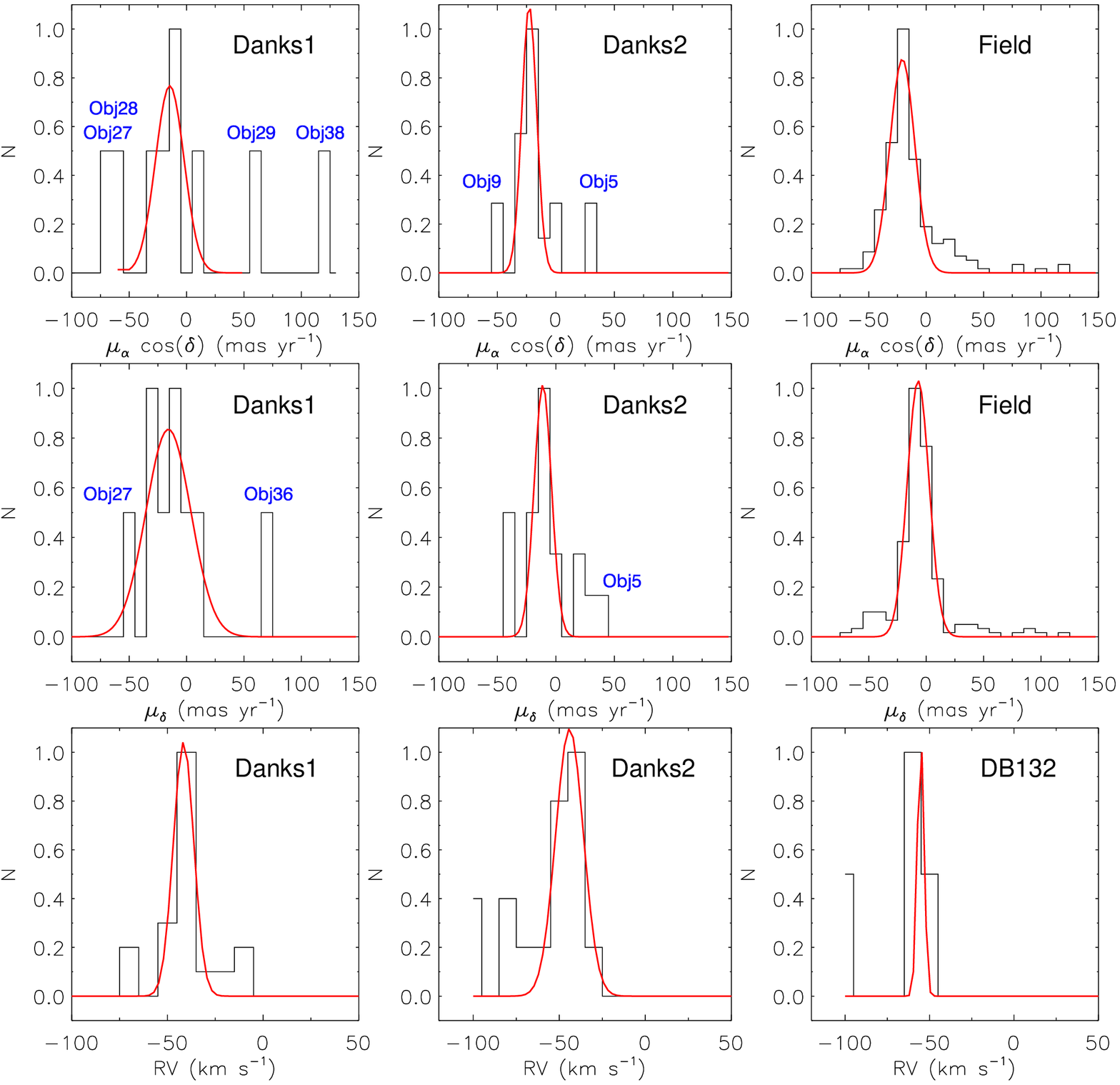}
   \includegraphics[width=9.0cm]{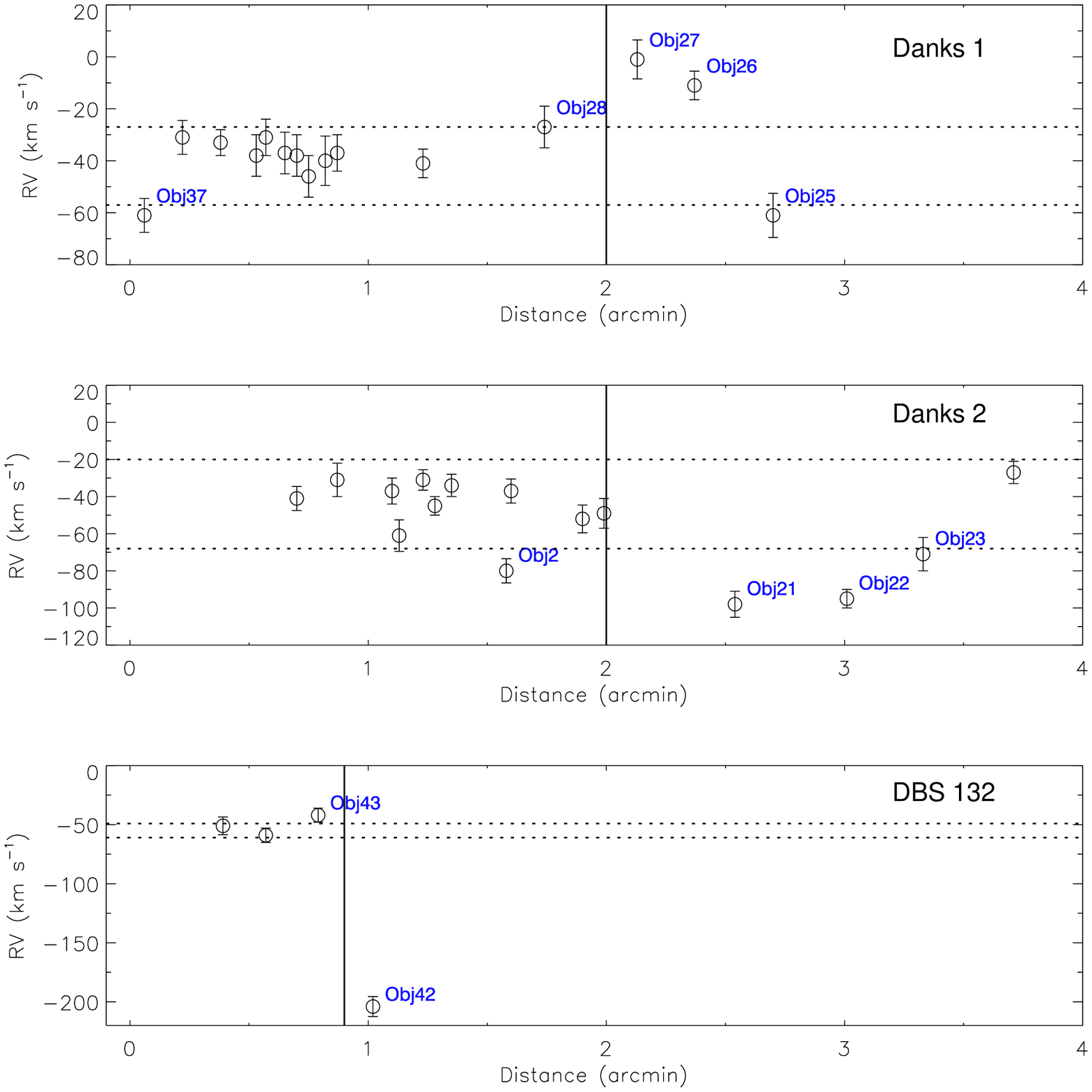}
   \caption{Left:Top and middle panel: Histograms and proper motion distributions of $\mu_{\alpha}\cos\delta$ and $\mu_{\delta}$ of
   Danks1, Danks2 and comparison field. Left: Bottom panel: RV distributions of Danks1, Danks2 and DB132. The solid lines represent the best Gaussian fit, outliers are labelled. Right: Radial velocities for our spectroscopic targets as a function of distance from the Danks\,1 (top), Danks\,2 (middle) and DBS\,132 (bottom) cluster centres. The horizontal lines represent the dispersion of the Gaussian fit (see text). The adopted cluster diameter is shown by the vertical line. The error bars represent the random error in determining the RV for each star.}
              \label{FigPMHist}%
\end{figure*}

We define the angular size of the clusters as the radius where the RDP meets the level of the field stars. An angular size of $1.5\pm0.5$~arcmin is found for both Danks\,1 and Danks\,2. \new{We also estimate S$_0=308\pm54$ stars/arcmin$^2$ and r$_{\rm{C}}=0.21\pm0.03$ for Danks\,1, and S$_0=267\pm98$ stars/arcmin$^2$ and r$_{\rm{C}}=0.32\pm0.06$ for Danks\,2}. RCW\,79 does not appear to follow a typical cluster profile. This could be due to artefacts affecting the image, since bright saturated stars leave big holes in the catalogue where the star count is artificially deficient. On the other hand, King profiles are meant to describe relaxed globular clusters, not young open clusters. As such it is not surprising that RCW\,79 shows a quite irregular form. Other profiles were also used, but with meaningless results. Still, it shows a stellar density excess at radius R~$<0.1$~arcmin that appears to reach R~$\sim 1$~arcmin, hence we can use this value as our estimate of the angular size. As for DB132, the stellar RDP is flat up to R~$\sim 3$~arcmin and so is not typical of a cluster.

\subsection{Memberships of the spectral targets and RVs}\label{mem}

The spectroscopic targets were selected from statistically decontaminated CMDs (Bonatto \& Bica \cite{Bon10}). However, as is well known, statistical decontamination methods are unable to perfectly determine cluster membership. As a result, we inevitably obtained some spectra of field stars. Cluster membership could be determined more accurately if proper motions were available. However, at present, the VVV database contains only two years worth of observations and so it is not yet possible to estimate precise proper motions from VVV data.

Therefore, we built the proper motion histograms using the absolute proper motions from the PPMXL Catalogue of Positions and Proper Motions on the ICRS (Roeser et al. \cite{Ro10}). At the distance of Danks\,1 and 2 (3--4 kpc, Davies  et al. \cite{Da11}, this work) the absolute proper motions are not very accurate (the mean error of the sample is $\sim$9~mas), but nevertheless can be used for some estimates (with 40--50\% accuracy). Unfortunately, only 9 of the spectroscopic targets in Danks1 and 18 of Danks 2 appear in the PPXML catalogue. The corresponding proper motion distribution histograms are shown in Figure~\ref{FigPMHist}, together with the proper motion distribution of a comparison field (RA~$=198.35231$ and DEC~$=-62.826599$). Using the Bonatto \& Bica (\cite{Bon11}) algorithm, we fitted the distributions with a Gaussian profile, defined by the velocity dispersion ($\sigma$) and the average velocity ($\mu$) of the stars (Table~\ref{table:3}). Stars outside the $5\sigma$ limit of the fit are considered field stars. Such a conservative limit is justified given the large uncertainties on the proper motions.

In addition, the RVs of the spectral targets can also be used to verify cluster membership. We measured the targets' RVs using the {\sc iraf} task {\it rvidlines}, which fits spectral lines to determine the wavelength shift with respect to specified rest wavelengths. In this procedure, we used all H{\sc i} and He{\sc i} lines available in the spectra (typically 7--10 lines per spectrum). We obtained an accuracy equivalent to roughly a tenth of a resolution element, typically $\sim$20~km~s$^{-1}$. As in the proper motion analysis, we fitted the RV distribution histogram with a Gaussian function. Then, we plotted the RV as a function of distance from the cluster centres (Figure~\ref{FigPMHist}, bottom left and right panels). The dashed lines on Figure~\ref{FigPMHist} correspond to the $3\sigma$ interval, and any star outside those limits is considered a field star. Objs\, 27 and 28 from Danks\,1 can be immediately classified as field stars. Obj\,37 (also observed by Davies et al. \cite{Da11}, their number D1-12) has a peculiar RV, but its spectrum implies cluster membership. The RVs of Danks\,2 are much more homogeneous, without obvious outliers. For DBS\,132, Obj\,42 has a RV that is much higher than the dispersion interval, so it is probably a field star. One should note that this analysis is based on a single epoch observation and follow-up observations of other epochs would be needed to eliminate bias due to binarity, especially since the binary fraction among OB and WR stars is high ($f_{bin}>0.5$ for OB stars; Sana et al. \cite{Sa11} and references therein; and $f_{bin}\sim0.4$ for WR stars; Schnurr et al. \cite{Sc08} and references therein). The average RVs of each cluster are listed in Table~\ref{table:3}. These appear consistent with $v_{LSR}=-39.4$~km~s$^{-1}$ (see Davies et al. \cite{Da11}, Figure~7). Additionally, Danks\,1 and Danks\,2 exhibit the same RV, which indicates that they are binary clusters. The estimated mean RV for DBS\,132 is significantly larger than for the Danks clusters, but it is based on only 3 stars. Note that Martins et al. (\cite{Ma10}) performed a similar analysis of RCW\,79, as such it is not repeated here.

\begin{figure}
  \centering
   \includegraphics[width=9.cm]{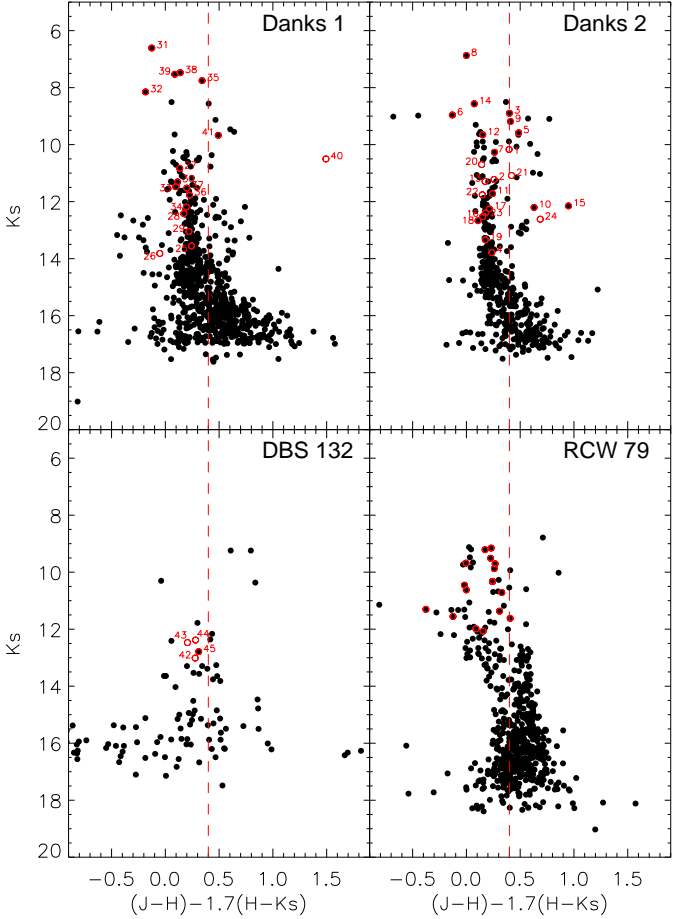}
   \caption{Reddening free colour of $(J-H)-1.70(H-K_s)$, also called $Q_{IR}$ plotted vs. K$_s$ magnitude. Red circles mark the stars for which spectra have been observed, and the number attached to them refers to the name of the star listed in Table\,\ref{table:2}. The red, vertical dashed line is placed at $Q_{IR}=0.4$ (see text for more details).}
              \label{FigQMD}%
\end{figure}

The final check that we performed was to analyse the position of the stars in the $J-K_s$ vs. $K_s$ CMD. Objs\,25, 34, 40 and 41 (Danks\,1) and Objs\,1, 3, 10, 14 and 21(Danks\,2) have peculiar positions in the diagram, far from the MS and the sequence of evolved stars. Of course, this could be due to differential reddening (see discussion in the next section). We attempted to reduce the effect of differential reddening by employing the reddening-free parameter $Q_{IR} =(J-H)-1.70(H-K_s)$, as defined by Negueruela et al. (\cite{Ne07}) for OB stars (see also Catelan et al. \cite{Ca11} for a list of several other reddening-free indices in the $ZYJHK$ system). We chose this parameter to avoid the intrinsic degeneracy between reddening and spectral type (and since we expect to find early OB stars in the majority of the clusters in our sample). Figure~\ref{FigQMD} shows this reddening free parameter vs. $K_s$ magnitude. Originally, Negueruela et al. (\cite{Ne11}) defined $Q_{IR} \leq$ 0.0-0.1 as a separating value for early-type stars. Subsequently, Ram{\'{\i}}rez Alegr{\'{\i}}a at al. (\cite{Ra12}) and Borissova et al. (\cite{Bo12}) defined the separation value for OB stars as $Q_{IR} \leq$ 0.3-0.4.  Indeed, plotting our objects in the reddening free diagrams, it would appear that Obj\,25 and 34 are affected by differential reddening, whilst Objs\,40 and 41 show infrared excess and are situated far from the standard OB sequence. Meanwhile, Objs\, 1, 3, 15, 21 and 24 of Danks\,2 occupy peculiar positions on the reddening free CMD. Obj\,3 was observed and classified by Davies et al. (\cite{Da11}, their number D2-2) as a yellow supergiant (F8-G3 I).\new{ It is difficult to determine, with any certainty, if this object is a cluster member. The radial velocity of the object is bigger than the mean value of the cluster(-61km/s), but the error is large $pm/$ 17km/s. Its proper motion values are also relatively far from the mean value. Most probably it is a field star, the conclusion Davies et al. (\cite{Da11}) also reach. If not, Danks\,2 will be another example of a cluster containing both supergiants and WR stars. However, accurate proper motion measurements are required for its status to be determined. Objs\, 1, 10, 21 and 24 are most probably field stars.}  All spectroscopically observed targets in DBS\,132 and RCW\,79 occupied the region expected of OB stars.  
 
\begin{table*}
\caption{Position, photometry and spectral parameters. Columns include the name of the star, the Davies et al. (\cite{Da11}, hereafter D12) identification number, the right ascension (RA), the declination (DEC), $V$ (taken from Baume et al. \cite{Ba09}), $J$, $H$ and $K_s$ photometry (in mag), the RV, the spectral type (determine from comparison with template spectra and $UBV$ photometry), extinction (A$_K$) and distance (d), respectively. Stars are ordered as a function of their (non-)membership.}             
\label{table:2}      
\centering                          
\begin{tabular}{lllccrrrrr@{$\pm$}lcccc}        
\hline\hline
\noalign{\smallskip}
&name  & D12 &         RA             &            DEC      &\multicolumn{1}{c}{$V$}&\multicolumn{1}{c}{$J$}&\multicolumn{1}{c}{$H$}&\multicolumn{1}{c}{$K_s$}& \multicolumn{2}{c}{RV} & {Sp. Typ.} & A$_K$ &    d\\
&      & name  &      (J2000)        &      (J2000)       &              &                &              &                & \multicolumn{2}{c}{(km s$^{-1}$)}    &                   &   (mag) & (kpc) \\
\noalign{\smallskip}
\hline
\noalign{\smallskip}
\multicolumn{4}{l}{\it Danks\,1 field stars:} &&&&&\multicolumn{2}{l}{}&&&\\
& Obj25 &       & 13 12 36.310 & -62 41 22.60  & 23.07 & 15.41  & 14.15 & 13.55 &  --61&17 & B9/A0V & 1.14$\pm0.05$ & 2.50\\
& Obj26 &       & 13 12 35.290 & -62 41 53.71  & 20.49 & 15.34  & 14.40 & 13.82 &  --11&11 & B9/A0V &  0.93$\pm0.03$ & 3.11\\
& Obj27 &       & 13 12 34.156 & -62 41 36.01  & 18.23 & 12.08  & 11.24 & 10.83 &  --1&15 & B2/3V & 0.82$\pm0.04$ & 2.15\\
& Obj28 &       & 13 12 32.760 & -62 42 07.61  & 20.43 & 13.99  & 12.94 & 12.42 &  --27&16 & B9/A0V & 0.96$\pm0.03$ & 1.61\\
& Obj29 &       & 13 12 30.240 & -62 41 29.40  & 20.85 & 14.46  & 13.49 & 13.05 &  --41&11 & B2/3V     & 0.92$\pm0.04$ & 5.71\\
& Obj40 &       & 13 12 23.234 & -62 42 01.00  & 20.23 & 12.22  & 10.59 & 10.50 &  --37&16 & O9/B0V & 1.15$\pm0.03$ & 3.56\\
& Obj41 & D1--8 & 13 12 22.879 & -62 41 48.83  & 18.64& 11.27   & 10.08 & 9.67   &  --31&14 & O4-6 & 1.08$\pm0.03$ & 4.04\\
\multicolumn{4}{l}{\it Danks\,1 members:} &&&&&\multicolumn{2}{c}{}&&&\\
& Obj30 &       & 13 12 29.299 & -62 42 07.12   & 21.66 & 12.67  & 11.77 & 11.31  &  --37&14 & B2/3V & 0.89$\pm0.06$ & 2.60\\
& Obj31 & D1--1 & 13 12 28.560 & -62 41 43.80   & 14.96 & 8.26    & 7.27   & 6.61     &  --46&16 & WNLh &  & \\
& Obj32 & D1--5 & 13 12 28.517 & -62 41 51.00   & 16.21 & 9.81    & 8.83   & 8.15     &  --38&16 & WNLh &  & \\
& Obj33 &       & 13 12 28.061 & -62 42 08.23   & 19.29 & 12.84  & 11.95 & 11.48  &  --31&14 & B2/3V & 0.89$\pm0.06$ & 2.81\\
& Obj34 &       & 13 12 27.459 & -62 41 19.27   & 21.41 & 13.99  & 12.78 & 12.19  &  --40&19 & B2/3V & 1.16$\pm0.07$ & 3.45\\
&       & D1--7 & 13 12 26.800 & -62 41 56.36   & 18.16 & 10.78  & 8.70   & 8.14    &  \multicolumn{2}{c}{}& O4-6 I & 1.74$\pm0.03$ & 2.42\\
&       & D1--11& 13 12 26.320 & -62 42 05.78   & 18.99 &             &            &             &  \multicolumn{2}{l}{} & O8-B3 V &  & \\
&       & D1--6 & 13 12 26.220 & -62 42 09.37   & 16.36 & 9.91    & 9.01   & 8.51     &  \multicolumn{2}{c}{}  & O6-8If & 0.99$\pm0.04$ & 4.06\\
& Obj35 & D1--4 & 13 12 26.218 & -62 41 57.90    & 16.95 & 9.61   & 8.32    & 7.76    &  \multicolumn{2}{c}{}  & O6-8If & 1.26$\pm0.03$ & 2.53\\
& Obj36 &       & 13 12 26.107 & -62 41 39.89    & 20.14 & 13.20 & 12.22 & 11.77  &  --33&10 & B2/3V & 0.93$\pm0.06$ & 3.16\\
&       & D1--9 & 13 12 26.020 & -62 42 15.59    & 19.01 & 11.34 & 10.28 & 9.69    &  \multicolumn{2}{c}{} & O4-6 & 1.11$\pm0.03$ & 4.01\\
& Obj37 & D1--12& 13 12 25.690 & -62 42 05.17    & 19.39  & 12.83 & 11.93 & 11.53 &  --61&13 & B2/3V & 0.86$\pm0.04$ & 2.92\\
& Obj38 & D1--2 & 13 12 25.006 & -62 42 00.21    & 17.40  & 9.422 & 8.15   & 7.48    &  --31&13 & WNLh &  & \\
&       & D1--10& 13 12 24.500 & -62 42 08.52  & 19.39  & 12.10 & 10.10 & 10.43  &  \multicolumn{2}{c}{} & O4-6 & 1.25$\pm0.03$ & 5.63\\
& Obj39 & D1--3 & 13 12 23.748 & -62 42 01.37   & 17.36  & 9.41   & 8.20   & 7.53     &  --38&16 & O7/8I  & 1.28$\pm0.03$ & 2.26\\
\multicolumn{13}{l}{}\\
\multicolumn{4}{l}{\it Danks\,2 field stars:} &&&&&\multicolumn{2}{c}{}&&&\\
& Obj1   &      & 13 13 11.340 & -62 41 10.68    & 12.25  & 10.62 & 10.19 &  10.17 & --41&12 & B9/A0V  & 0.28$\pm0.05$ & 0.78\\
& Obj2   &      & 13 13 00.840 & -62 39 54.66    & 17.92  & 12.32 & 11.53 &  11.22 & --80&13 & B3/4V  & 0.73$\pm0.06$ & 2.40\\
& Obj3   & D2-2 & 13 12 59.986 & -62 40 40.16     & 12.19  & 9.61   & 9.01    &  8.90   & --61&17 & F8-G1 &  & \\
& Obj10  &      & 13 12 56.112 & -62 40 53.90    & 16.41  & 14.42 & 12.79 & 12.20  & \multicolumn{2}{c}{} & B2/3V:   & 1.41$\pm0.06$ & 3.09\\
& Obj21 &       & 13 12 45.300 & -62 40 42.16    & 18.43  & 12.61 & 11.50 & 11.08  & --98&14 & G7/8I &  & \\
& Obj22 &       & 13 12 43.459 & -62 40 57.40    & 18.22  & 12.87 & 12.12  & 11.76 & --95&10 & B5/6V  & 0.72$\pm0.05$ & 2.21\\
& Obj23 &       & 13 12 42.250 & -62 41 11.42    & 19.80  & 13.65 & 12.79  & 12.40 & --71&18 & B5/6V  & 0.80$\pm0.07$ & 2.85\\
& Obj24 &       & 13 12 41.290 & -62 41 51.40    & 20.15  & 13.72 & 12.77  & 12.62 & --27&12 & B3/4V & 0.73$\pm0.07$ & 4.56\\
\multicolumn{4}{l}{\it Danks\,2 members:} &&&&&\multicolumn{2}{c}{}&&&\\
& Obj4   &      & 13 12 58.930 & -62 40 46.49     & 19.32  & 15.05 & 14.17 & 13.79  & --31&18 & B8/9V & 0.78$\pm0.04$ & 3.81\\
&        & D2--7 & 13 12 58.560 & -62 40 54.84    & 16.73  & 10.76 & 9.95    & 9.56    & \multicolumn{2}{c}{} & O4/6V  & 0.84$\pm0.05$ & 4.75\\
& Obj5   & D2--6 & 13 12 58.351 & -62 40 37.34     & 19.60  & 10.86 & 10.04  & 9.63    & \multicolumn{2}{c}{} & O8/9V  & 0.85$\pm0.04$ & 3.37\\
& Obj6   & D2--3 & 13 12 57.739 & -62 40 59.91     & 17.25  & 10.83 & 9.83    & 8.98    & \multicolumn{2}{c}{}  & WC7-8  &  & \\
& Obj7   &       & 13 12 57.737 & -62 40 40.13     & 17.58  & 11.50 & 10.63  & 10.27 &  \multicolumn{2}{c}{} & O9/B0V:  & 0.85$\pm0.03$ & 3.67\\
&        & D2--9 & 13 12 57.083 & -62 40 00.49    &              & 15.19 & 13.42  & 12.64 & \multicolumn{2}{c}{} & O6-8 &  & \\
& Obj8   & D2--1 & 13 12 56.419 & -62 40 28.43     & 14.79  & 8.30   & 7.40     & 6.87   &  \multicolumn{2}{c}{}  & O8-B3I  & 1.00$\pm0.04$ & 1.90\\
& Obj9   & D2--4 & 13 12 56.266 & -62 40 51.24     & 16.41  & 10.47 & 9.51    & 9.19   &  \multicolumn{2}{c}{}  & O6/7 V   & 0.89$\pm0.05$ &  3.53\\
& Obj11  &        & 13 12 54.871 & -62 41 03.96    & 18.73   & 12.89 & 12.06 & 11.72  & \multicolumn{2}{c}{} & B2/3V: & 0.77$\pm0.06$ & 3.32\\
& Obj12  & D2--5 & 13 12 54.485 & -62 41 05.01    & 16.62   & 10.80 & 10.02 & 9.65    & \multicolumn{2}{c}{} & O7/8V  & 0.81$\pm0.03$ & 4.09\\
&        & D2--8 & 13 12 54.370 & -62 40 45.48     & 16.80   & 11.14 & 10.34 & 9.92   &  \multicolumn{2}{c}{}  & O6-8V  & 0.84$\pm0.05$ & 3.87\\
& Obj13 &        & 13 12 52.730 & -62 40 54.51    & 17.93   & 12.41 & 11.64 & 11.29 &  --41&13 & B0/1V  & 0.77$\pm0.07$ & 4.91\\
& Obj14 &        & 13 12 51.773 & -62 39 55.96    & 18.75   & 11.91 & 9.78    & 8.57   & --31&11   & O7/8I & 2.16$\pm0.03$ & 2.43\\
& Obj15 &        & 13 12 51.221 & -62 41 02.38    & 19.43   & 13.17 & 12.18 & 12.15 & --37&14 & B3/4V & 0.68$\pm0.06$ & 3.76\\
& Obj16 &        & 13 12 50.541 & -62 40 22.86    & 19.54   & 13.72 & 12.91 & 12.53 & --45&10 & B5/6V  & 0.76$\pm0.07$ & 3.08\\
& Obj17 &        & 13 12 50.160 & -62 41 00.96    & 18.85   & 13.41 & 12.60 & 12.26 & --34&12 & B5/6V & 0.74$\pm0.07$ & 2.75\\
& Obj18 &        & 13 12 49.159 & -62 41 00.20    & 19.39   & 13.82 & 13.05 & 12.67 & --37&13 & B5/6V & 0.74$\pm0.07$ & 3.32\\
& Obj19 &        & 13 12 48.130 & -62 41 13.36    & 19.88   & 14.44 & 13.68 & 13.34 & --52&15 & B7/8V  & 0.69$\pm0.13$ & 3.71\\
& Obj20 &        & 13 12 47.520 & -62 40 54.48    & 17.40   & 11.82 & 11.06 & 10.60 & --49&16 & B0/1V  & 0.77$\pm0.07$ & 3.74\\
\multicolumn{13}{l}{}\\
\multicolumn{4}{l}{\it DBS2003\,132 field star:} &&&&&\multicolumn{2}{c}{}&&&\\
& Obj42 &       & 13 12 22.270 & -62 42 38.28     & 22.05 & 14.70 & 13.53 & 13.01 &  --204&17 & B7/8V & 1.04$\pm0.13$ & 2.49\\
\multicolumn{4}{l}{\it DBS2003\,132 members:} &&&&&\multicolumn{2}{c}{}&&&\\
& Obj43 &       & 13 12 21.004 & -62 41 48.41     & 20.78 & 13.98 & 12.95 & 12.47 &  --42&12 & B5/6V & 0.96$\pm0.07$ & 2.87\\
& Obj44 &       & 13 12 19.871 & -62 42 25.23     & 23.02 & 14.41 & 13.03 & 12.38 &  --51&15 & B3/4V & 1.30$\pm0.06$ & 3.35\\
& Obj45 &       & 13 12 16.202 & -62 42 10.20     & 21.06 & 15.39 & 13.64 & 12.79 &  --59&12 & B3/4V & 1.64$\pm0.06$ & 3.44\\
\noalign{\smallskip}
\hline
\end{tabular}
\end{table*}

The reddening free CMDs can also be used to separate the MS from the pre-main sequence (PMS) stars. In all cases, except for DBS\,132 the PMS population is well separated, and can be defined as stars with $Q_{IR} >$ 0.4 and  $K_s > 14$; $K_s > 15$; $K_s > 14$ and  $K_s > 13$ mag for Danks\,1,  Danks\,2, DBS\,132, and  for RCW\,79, respectively. This separation was used during isochrone fitting and mass calculation. 
 
\subsection{Extinction}

The new spectroscopic classifications \new{and infrared observations permit the extinction and reddening law to be evaluated for the lines-of-sight towards each star studied. The reddening law is known to vary throughout the Galaxy. For example, Turner (\cite{Tu11}) noted that dust in the direction of the $\eta$ Car complex follows an anomalous reddening law.} A variable extinction analysis (e.g. Turner \cite{Tu76}, Majaess et al. \cite{Maj11}) may be employed to evaluate both the ratio of total-to-selective-extinction (e.g., $A_{K_s}/E(J-{K_s})$) for the line-of-sight and the distance to Danks\,1 and Danks\,2. Substantial samples are required for the analysis, and thus the clusters were assumed to lie at the same distance. That assumption is supported by the new RVs derived here and the results of Davies et al. (\cite{Da11}). From the expression for computing the distance to a star we get:

\begin{eqnarray}\label{EqAk}
\nonumber
K_s-M_{K_s}-R_{K_s} \times E(J-{K_s}) =5\times \log{d}-5, \\
\end{eqnarray}

\noindent where we define $R_{K_s}=A_{K_s}/E(J-K_s)$. The equation simplifies for a cluster of stars at a common distance. Rearranging Equation\,(\ref{EqAk}) yields:

\begin{eqnarray}
\nonumber
(K_s-M_{K_s}) =R_{K_s} \times E(J-{K_s}) + \rm{const}. \\
\end{eqnarray}

$R_{K_s}$ and $5\times \log{d}-5$ were estimated by comparing $K_s-M_{K_s}$ and $E(J-{K_s})$. $M_{K_s}$ and $E(J-{K_s})$ were inferred from the new spectroscopic observations by adopting intrinsic parameters from Schmidt-Kaler (\cite{Sc82}) and Koornneef (\cite{Ko83}). A least squares fit to the data yields $d=3.6\pm0.5$~kpc and $R_{K_s}=0.60\pm0.24$. The results obtained favour the larger distance to Danks\,1 and Danks\,2 proposed by Davies et al. (\cite{Da11}). This topic is elaborated upon in the next section.

Subsequently, $A_{K_s}$ was estimated for each star, using the value of $R_{K_s}$ derived above. The averages derived from all cluster members are $A_{K_s} = 1.11 \pm 0.24$~mag for Danks\,1 and $A_{K_s} = 0.88 \pm 0.34$~mag for Danks\,2. The results agree with those determined by Davies et al. (\cite{Da11}), within the uncertainties. For DBS\,132 we calculated $A_{K_s} = 1.30\pm0.34$~mag. We did not obtain spectra of RCW\,79's members, but adopting the spectral classification of the stars observed by Martins et al. (\cite{Ma10}), we get $A_{K_s} = 0.56\pm0.04$~mag. The errors represent only the standard deviation from the mean value and can be relatively large in some cases, suggesting differential extinction. In Danks\,2, only one star has a particularly large reddening value, namely Obj\,14, for which of $A_{K_s}$ = 2.16~mag. Though we note that this star exhibits peculiar behaviour. Excluding it from the calculation we get $A_{K_s}$ = 0.80$\pm0.08$~mag for Danks\,2. The reddening of DBS\,132 is calculated using only 3 spectroscopically observed stars, but the colour spread of DBS\,132's cluster CMDs (see below) is $\sim 1$~mag. This is much larger than the typical photometric errors, suggesting significant differential extinction. The colour spread is $\sim 0.4$~mag in Danks\,1 and $\sim 0.3$~mag  in RCW\,79, much smaller than in DBS\,132, but still significant.

\new{The mean colour-excess inferred for stars in Danks 2 is: $E(J-K_s) =1.21\pm 0.02(\sigma_{\bar{x}}) \pm 0.09 (\sigma)$~mag, $E(J-H) =0.85\pm 0.02(\sigma_{\bar{x}}) \pm 0.07 (\sigma)$~mag and $E(B-V) =2.40\pm 0.06(\sigma_{\bar{x}}) \pm 0.24 (\sigma)$~mag, where $\sigma$ is the standard deviation and $\sigma_{\bar{x}}$ is the standard error, i.e. $\sigma$ divided by the square root of the sample size. The aforementioned mean reddenings were inferred from the intrinsic $BVJH$ colours of Turner (\cite{Tu89}) and Strai\v zys \& Lazauskait\.e (\cite{St09}), so as to provide an independent check on the extinction estimates determined from the intrinsic parameters of Schmidt-Kaler \cite{Sc82} and Koornneef \cite{Ko83} in the preceding paragraph. The mean colour excess inferred from the spectroscopic observations for stars in Danks 1 is: $E(J-K_s) =1.63\pm 0.05(\sigma_{\bar{x}}) \pm 0.21 (\sigma )$, $E(J-H) =1.07\pm 0.04(\sigma_{\bar{x}}) \pm 0.15 (\sigma )$ and $E(B-V) =2.68\pm 0.06(\sigma_{\bar{x}}) \pm 0.27 (\sigma )$. Dust emission observed in WISE images of the field support the finding that Danks\,1 suffers additional obscuration relative to Danks\,2.

Baume et al. (\cite{Ba09}) discovered that the canonical reddening law does not characterize dust associated with the target clusters. The ratio $E(J-H)/E(H-K_s)$ may be inferred from all stars observed along the line of sight. A median (after making a $1.5\sigma_{\bar{x}}$ clip) yields $E(J-H)/E(H-K_s)=2.20\pm0.06(\sigma_{\bar{x}}) \pm 0.37 (\sigma )$, which is far in excess of the various values for the inner Galaxy listed in Table~1 of Strai\v zys \& Lazauskait\.e (\cite{St08})'s work. It is also larger than the values tabulated in Majaess et al. (\cite{Maj11}) and Majaess et al. (\cite{Maj12}). The results imply that the line-of-sight $\ell\sim305\degr$ exhibits an anomalous reddening law, thus \new{corroborating the prior findings of} Baume et al. (\cite{Ba09}). The average $JHK_s$ reddening law for the inner Galaxy is $E(J-H)/E(H-K_s) \sim2.0$ (Strai\v zys \& Lazauskait\.e \cite{St09}), and Majaess et al. (\cite{Maj11}) noted that $E(J-H)/E(H-K_s)\sim1.94$ appears to describe dust in the field of the classical Cepheid TW Nor. The region encompassing Danks\,1 and Danks\,2 was not surveyed by Strai\v zys \& Lazauskait\.e (\cite{St09}) in their analysis of the reddening law throughout the inner Galaxy. The reddening law determined was subsequently employed to interpret the $JHK_s$ CMDs. An evaluation of the the ratio of total to selective extinction and reddening law in the optical $(E(U-B)/E(B-V)$, $R_V)$ did not yield viable results, presumably since the uncertainties are magnified owing to the added temperature sensitivity of optical passbands and uncertainties tied to acquiring solid $U$-band photometry (a challenge).}

\begin{figure}
  \centering
   \includegraphics[width=9.cm]{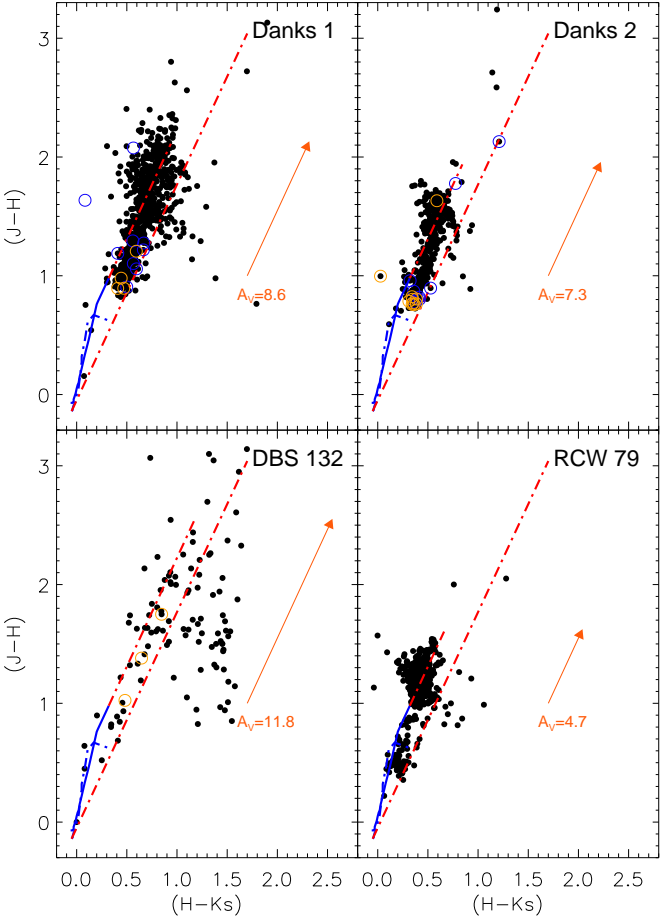}
   \caption{$(J-H)$ vs. $(H-K_s)$ colour-colour diagram. The continuous lines represents the sequence of the zero-reddening stars of luminosity class I and class V (Koornneef \cite{Ko83}). The reddening vector is overplotted and the dotted lines are parallel to the standard reddening vector. O and B stars are marked with blue and orange circles, respectively.}
              \label{FigCCD}%
\end{figure}

A field star decontaminated $J-H$ vs. $H-K_s$ CCD is presented in Figure~\ref{FigCCD}. O and B stars are marked with blue and orange circles, respectively. The intrinsic colours of the MS stars and giant branch (Koornneef \cite{Ko83}) are plotted using blue dashed lines. The reddening vector determined above is plotted in orange. Sources located to the right and below the reddening line may have excess emission in the near infrared (IR-excess sources) and/or may be PMS stars.

In this paper we will use $A_{K_s} = 1.11\pm0.24$, $A_{K_s} = 0.88\pm0.34$~mag, $A_{K_s} = 1.30\pm0.34$~mag and $A_{K_s} = 0.56\pm0.04$~mag for Danks\,1, Danks\,2, DBS\,132 and RCW\,79, respectively. The results are tied to the intrinsic parameters of Schmidt-Kaler \cite{Sc82} and Koornneef \cite{Ko83}, since $R_{K_s}$ was derived using these studies. Strai\v zys \& Lazauskait\.e \cite{St09} supply intrinsic colours only. The parameters derived are consistent, within the uncertainties, using either dataset.

\begin{table*}
\begin{center}
\caption{Proper motion and RV components}
\label{table:3}
\begin{tabular}{lrrrrrrr}
\hline 
\multicolumn{1}{l}{Name} &
\multicolumn{1}{l}{N} &
\multicolumn{1}{l}{Mean ($\mu_{\alpha}\cos\delta$)} &
\multicolumn{1}{l}{$\sigma_{\mu_{\alpha}\cos\delta}$} & 
\multicolumn{1}{l}{Mean ($\mu_{\delta}$)} &
\multicolumn{1}{l}{$\sigma_{\mu_{\delta}}$} &
\multicolumn{1}{l}{Mean (RV)} &
\multicolumn{1}{l}{$\sigma_{RV}$} \\
\multicolumn{1}{l}{} &
\multicolumn{1}{l}{stars} &
\multicolumn{1}{c}{mas yr$^{-1}$} &
\multicolumn{1}{c}{mas yr$^{-1}$} &
\multicolumn{1}{c}{mas yr$^{-1}$} &
\multicolumn{1}{c}{mas yr$^{-1}$} &
\multicolumn{1}{c}{km s$^{-1}$} &
\multicolumn{1}{c}{km s$^{-1}$} \\
\hline
Field   & 194 & --21$\pm1$ &  11$\pm1$ & --7$\pm1$ &  10$\pm1$ &$--$ & $-$ \\
Danks1  & 9   & --28$\pm2$ &  12$\pm3$ & --16$\pm4$&  19$\pm4$ & --42$\pm1$ & 5.0$\pm0.5$ \\
Danks2  & 18  & --23$\pm1$ &  6$\pm1$  & --11$\pm1$&  7$\pm1$  & --44$\pm1$ & 8.0$\pm1.0$ \\
DB132   & 5   & $-$       & $-$       & $-$      & $-$       & -55$\pm5$ &  2.0$\pm1.0$ \\
\hline
\end{tabular}
\end{center}
\end{table*}

\subsection{Distances}

The distances to the stars for which we have obtained spectra were calculated and are presented in Table\,\ref{table:2}. The extinction and $M_K$ values determined in the previous section were used to estimate the distance modulus. We then estimated the distance modulus to each cluster by taking the mean of the estimates for its members, using only the members for which we are confident of the spectral type. We get distances of: $(m-M)_{0}=12.7\pm0.6$~mag \new{(3.5$\pm1.0$~kpc, 8 stars, dwarfs only, namely: Objs\,30, 33, 34, 36, 37, 40, 4, D1-10) for Danks\,1; $(m-M)_{0}=12.8\pm0.5$~mag (3.7$\pm0.5$~kpc, 15 stars, dwarfs only, namely: Objs\,4, 5, 7, 9, 11, 12, 13, 15, 16, 17, 18, 19, 20, D2-8, D2-7) for Danks\,2; $(m-M)_{0}=12.5\pm0.2$~mag (3.2$\pm0.3$~kpc, 3 stars)} for DBS\,132. The quoted errors are the standard deviation of the individual stellar distance estimates. Thus we \new{support} Davies et al. (\cite{Da11})'s conclusion that Danks\,1 and Danks\,2 (and, now, also DBS\,132) have the same distance, to within the uncertainties. \newII{These distances also corroborate even earlier values obtained by Bica et al. (\cite{Bi04}).} As for RCW\,79, we adopt  the spectral classification of the stars observed by Martins et al. (\cite{Ma10}) to determine the distance, and get $(m-M)_{0}=13.4\pm0.4$~mag (4.8$\pm0.8$~kpc).

\subsection{Ages}

\begin{figure*}
  \centering
   \includegraphics[width=9.cm]{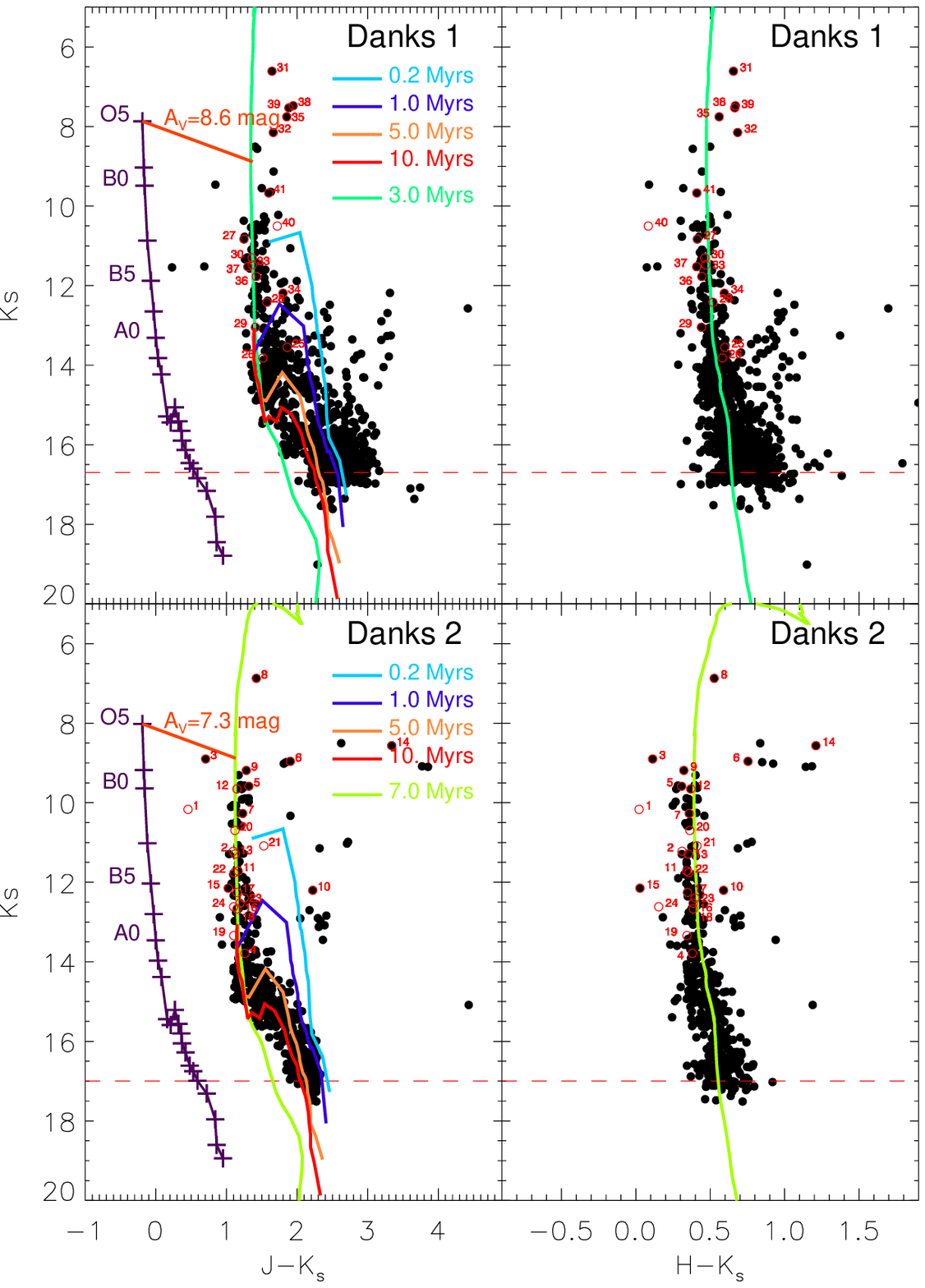}
   \includegraphics[width=9.cm]{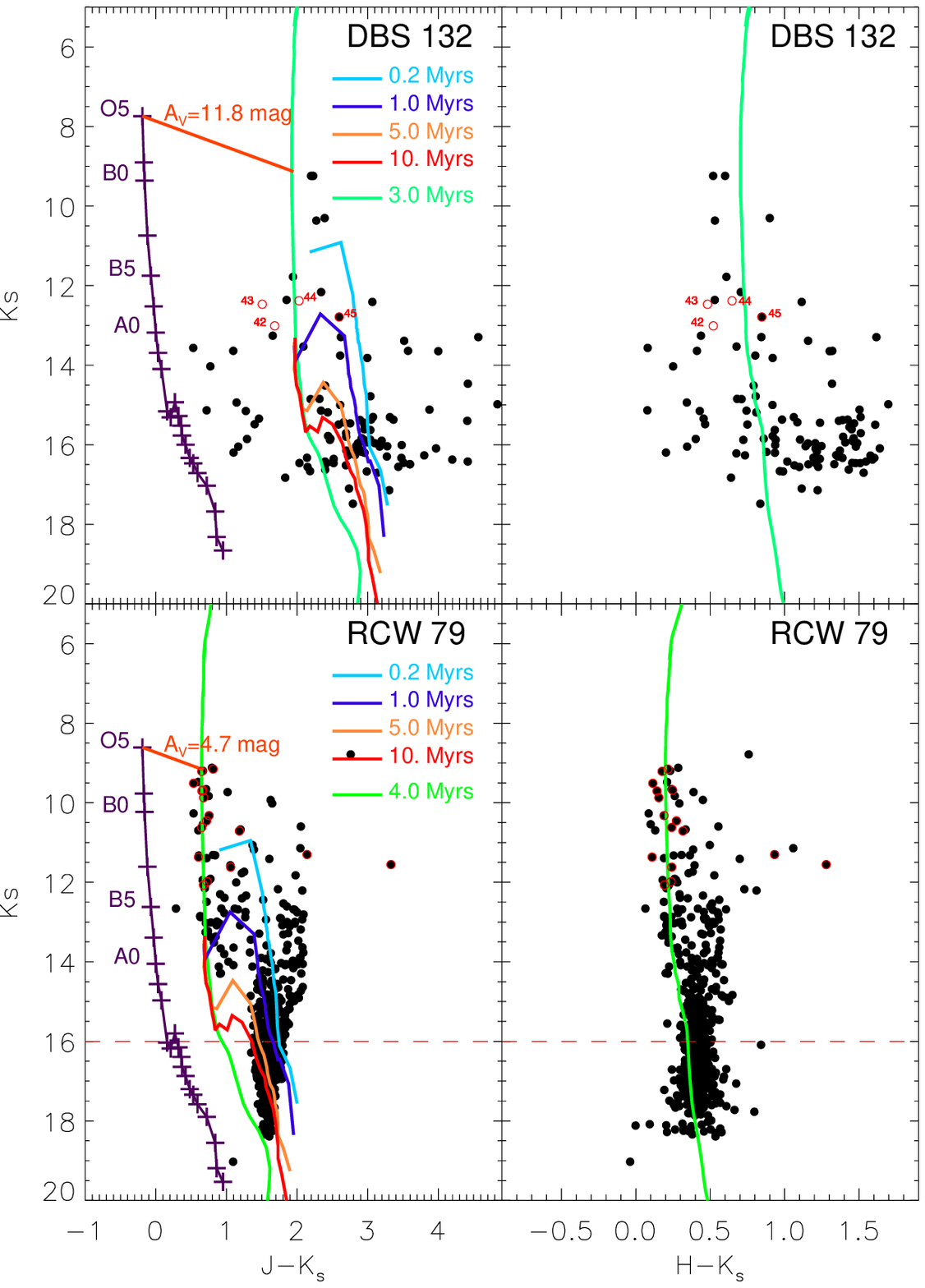}
   \caption{ The colour-magnitude diagrams for the four studied clusters. The Schmidt-Kaler (\cite{Sc82}) sequence is shown in the $(J-K_s)$ vs. $K_s$ diagram. Red circles mark the stars for which spectra have been observed, and the number attached to them refers to the name of the star listed in Table\,\ref{table:2}. The red dashed lines represent 50\% completeness limit. \newII{The green line is the main-sequence model (the exact colour changes with the age used), whereas the others (light blue, dark blue, orange and read) are PMS models.} See text for more details.}
              \label{FigCMD}%
\end{figure*}

\begin{figure}
  \centering
   \includegraphics[width=7.cm]{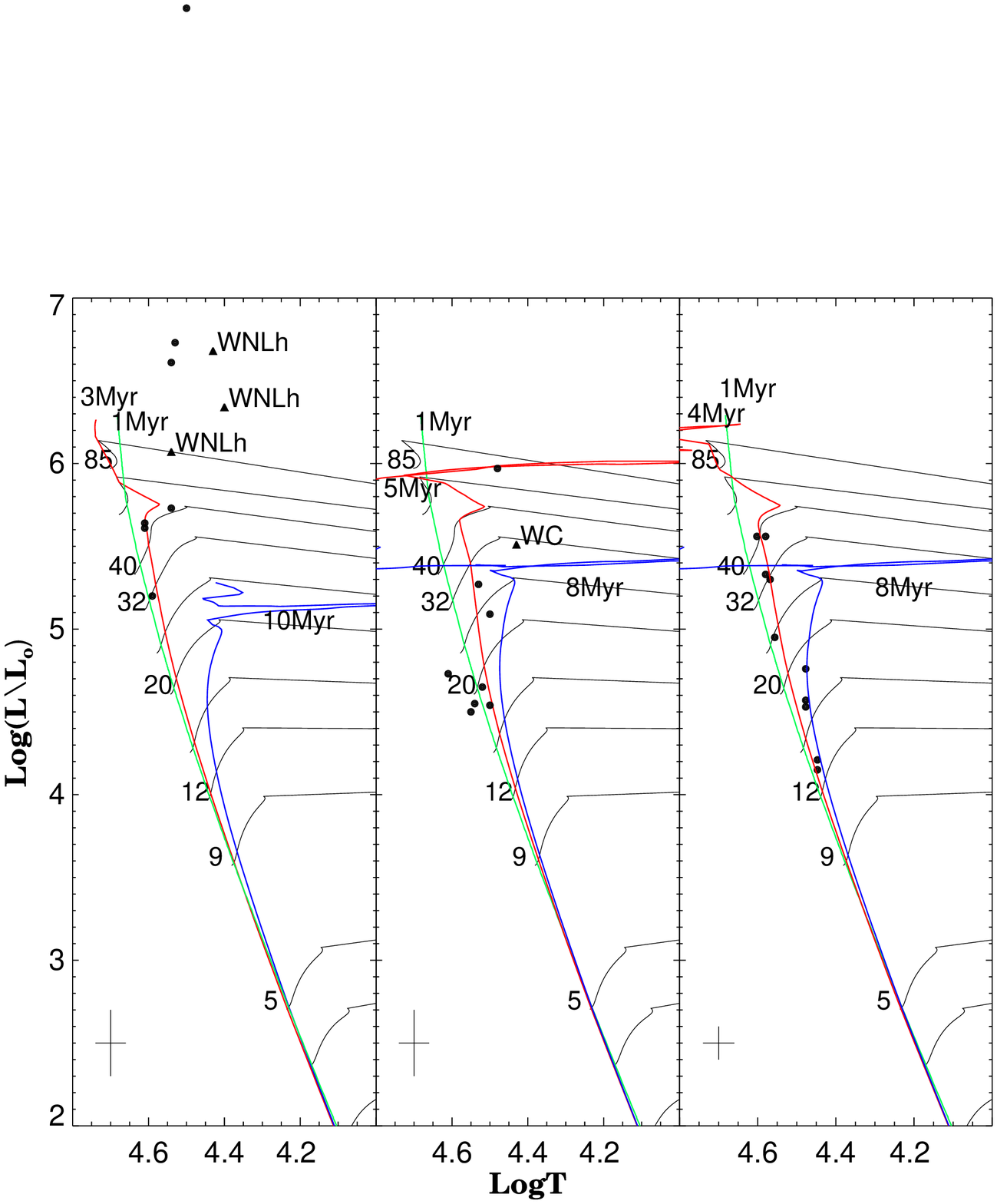}
   \caption{Hertzsprung-Russell diagrams for the Danks1, Danks2 and RCW79. Circles represent the early-type O stars, triangles the WR stars. The isochrones and stellar evolutionary tracks are from Ekstr{\"o}m et al. (\cite{Ek12}). 
}
              \label{Fighr_all}%
\end{figure}

The ages were estimated by fitting the observed CMDs with solar-metallicity \new{Geneva isochrones (Lejeune \& Schaerer \cite{Le01}) and PMS isochrones (Seiss et al. \cite{Se00})}. Starting with the isochrones set to the previously determined distance modulus and reddening, we apply shifts in magnitude and colour until the fitting statistics reach a minimum value (i.e. difference in magnitude and colour of the stars from the isochrone should be minimal) \new{for both the main sequence and PMS isochrones. We estimate ages of 1-5~Myr for Danks\,1, 4-7~Myr for Danks\,2, 1-3~Myr for DBS\,132 and 2-4~Myr for RCW\,79. Figure\,\ref{FigCMD} shows the isochrone fits superimposed on the decontaminated CMDs for all clusters.  

However, as can be seen in the Figure\,\ref{FigCMD}, for such young star clusters the main sequence isochrones are almost vertical lines in the near-infrared. Therefore, it is hard to determine the precise age, even using the PMS isochrone set. For this reason and following Liermann et al. (\cite{Li12}), we constructed HR diagrams using only the most luminous stars in Danks 1, Danks2 and RCW79 (Figure\,\ref{Fighr_all}). The effective temperatures for the early-type OB type stars were determined from the spectral types of the stars according to the spectral--type--temperature calibration of Martins et al. (\cite{Ma05}). The bolometric corrections are calculated using equation 2 of Liermann et al. (\cite{Li12}) and their correction to BC$_K$. The bolometric corrections for WR stars are taken from Martins et al. (\cite{Ma08}).  Absolute stellar magnitudes were derived from the $K_S$ magnitudes given in Table\,\ref{table:2}, with extinction and distance derived from their spectral types. 

We then compared the distribution of the stars with the latest isochrones and stellar tracks (with rotation) taken from  Ekstr{\"o}m et al. (\cite{Ek12}). The temperatures and luminosities of stars in RCW79 were taken directly from Martins et al. (\cite{Ma10}). Here we repeated their Fig.~7, though now the same isochrones and stellar tracks system as with the other clusters. As can be seen, most of the stars are concentrated around  3, 5 and 4 Myr isochrones for Danks1, Danks2 and RCW79 respectively. In comparison to the Davies et al. (\cite{Da11}) age determination, we derived slightly older ages for Danks1 and Danks2. However, taking in to account the large uncertainties on the determined ages, we consider our results in good agreement. Martins at al (\cite{Ma10}) also derived a younger age for RCW\,79 (2.3$\pm$0.5Myr), but again, this is consistent with our determination.}

\subsection{Initial mass function and cluster masses}

By looking at the clusters' CMDs it was often obvious which stars were not genuine cluster members or were not MS objects. Some of these spurious objects are by-products of our decontamination method (see section \ref{decont}). There were also some Wolf-Rayet stars that needed to be removed. The LF was converted to the Initial Mass Function (IMF) using the \new{Geneva isochrones, assuming solar metallicity and ages of 3, 5, 4 and 2 Myrs for Danks\,1, Danks\,2, RCW\,79 and DBS\,132 respectively}. For each cluster we established the MS turn-on point and assumed everything fainter than this to be PMS. This point was found to be at $K_s\sim$14 for Danks\,1, Danks\,2 and DB\,132, and $K_s\sim$13 for RCW79. 

Our calculated slopes of \new{$\Gamma$=$-$1.43$\pm$0.17 for Danks\,1 and $\Gamma$=$-$1.23$\pm$0.22 for Danks\,2,} are consistent with the Salpeter slope ($\Gamma$=$-$1.35), to within 1$\sigma$, \new{as well as with Davies et al. (\cite{Da11}) results}. These slopes were used to extrapolate the masses of the observed MS stars to provide a total mass for the cluster. For extrapolation of masses below $<$0.5\,M$_\odot$ we used the Kroupa value of $\Gamma$=$-$0.3. This gave a total mass of \new{\newII{7900$^{+1400}_{-1050}$\,M$_{\odot}$ for Danks\,1 and 2900$^{+850}_{-550}$\,M$_{\odot}$ for Danks\,2.} The results are in very good agreement with those found in Davies et al. (\cite{Da11}): $8000\pm1500$\,M$_\odot$ for Danks\,1 and $3000\pm800$\,M$_\odot$ Danks\,2.}

RCW\,79 has a slope \new{of $\Gamma$=$-$1.05$\pm 0.28$, with a mass of \newII{3000$^{+950}_{-450}$M$_{\odot}$.}} DB\,132 did not contain enough stars to reliably estimate a slope, nor completeness for each magnitude band. However, using the same methodology as demonstrated above, we suggest a minimum mass of $\sim$400\,M$_{\odot}$.

It should be noted that the stellar mass function is not the same as the IMF, despite the young age of the clusters. Dynamical evolution of the cluster will result in the loss of stars through the equipartition of energy, leading to a higher average star-mass and an increase in the binarity proportion. This will affect the slope of the stellar mass function, flattening it. This process is dependant on both cluster mass and lifetime, particularly for those with $<$10$^4$\,M$_{\odot}$ and life-times of a few 100\,Myr (Kroupa et al. \cite{Kr11}). Therefore, the clusters' dynamical evolution must be considered.

\begin{table}[!h]
\caption{Physical parameters of the clusters}
\label{results}
\begin{minipage}{180mm}
\begin{tabular}{lcccc}
\hline
Cluster    & Slope & Mass (PMS) & Mass (MS) &                Total Mass \\
           &       &   M$_\odot$ & M$_\odot$   &  M$_\odot$  \\  
\hline
\vspace{.1cm}
Danks\,1  & $-$1.43$\pm 0.17$  &     900$^{+50}_{-30}$       &      7000$^{+1450}_{-1000}$      &   7900$^{+1400}_{-1050}$ \\  
\vspace{.1cm}
Danks\,2  & $-$1.23$\pm 0.22$  &     600$^{+50}_{-50}$       &      2300$^{+800}_{-500}$       &   2900$^{+850}_{-550}$  \\   
\vspace{.1cm}
RCW\,79   & $-$1.05$\pm 0.28$  &     500$^{+200}_{-100}$      &      2500$^{+750}_{-300}$       &   3000$^{+950}_{-450}$  \\  
\vspace{.1cm}
DB\,132   & $-$0.97$\pm 0.39$  &   $>$60       &    $>$265         & $>$345    \\  
\hline
\end{tabular}
\end{minipage}
\end{table}

\section{Search for variable stars}\label{var}

The VVV survey provides many epochs of observations in the $K_s$-band. So far, now that the second year of the survey is completed, a small fraction of the total number of expected epochs have been obtained in the Galactic disk. This could be enough to look for variable candidates and perhaps obtain some periodicities. An observation log of the $K_s$ observations of the two VVV fields which contain the clusters studied in this paper is presented in Table\,\ref{table:1}.

\begin{figure}
  \centering
   \includegraphics[width=8cm]{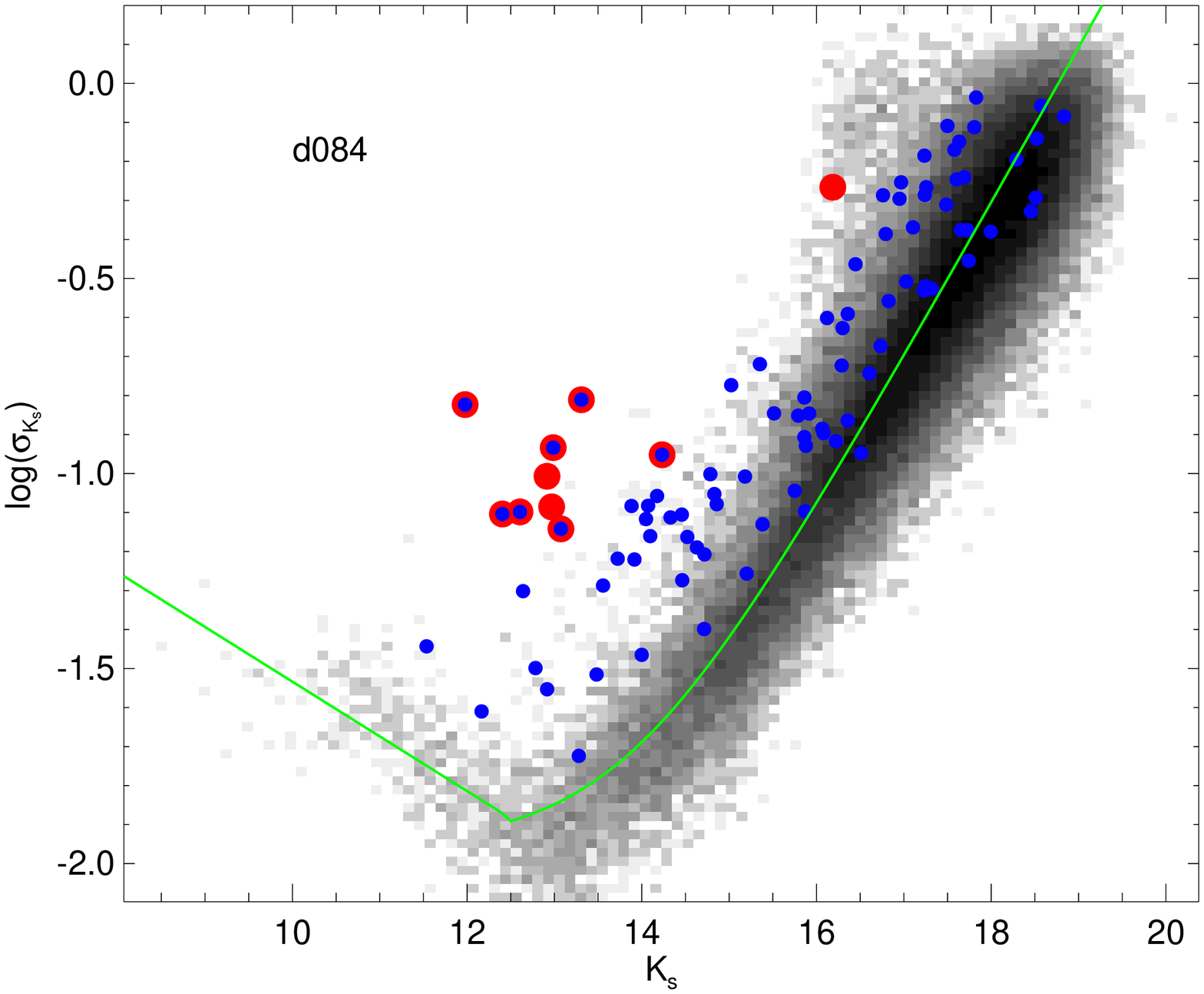}
   \includegraphics[width=8cm]{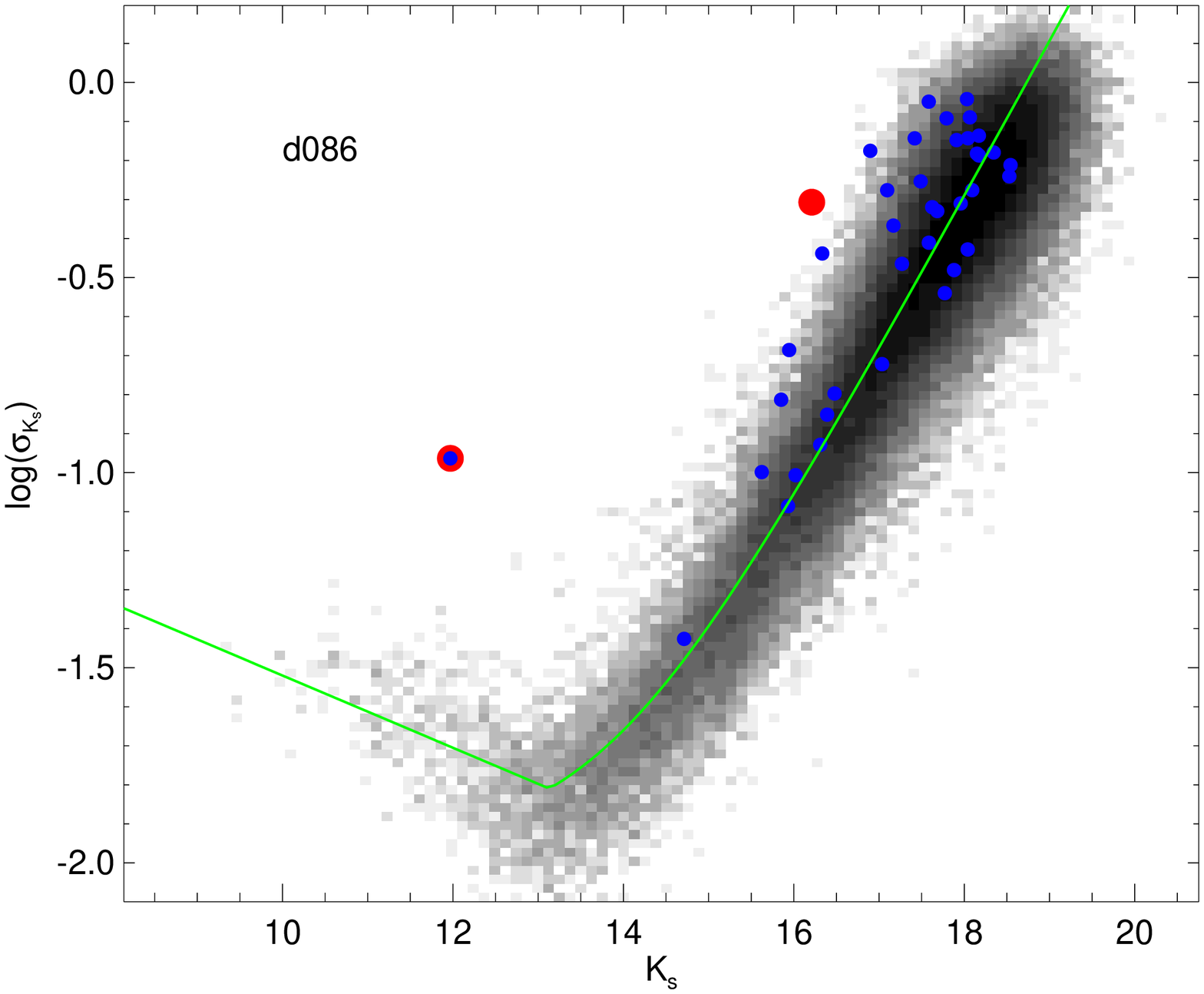}
      \caption{Standard deviation ($\sigma_{K_s}$) of the $K_s$-band light-curve extracted from the VVV fields d084 and d086. The fit of a theoretical relation between the $K_s$ and the photometric accuracy is plotted in solid, green line. The variable candidates among the stars in the radius of the four studied clusters are shown. Big, red circles are stars having a variability amplitude higher than $5\sigma_{K_s}$, while small, blue circles are stars with $\sigma_{K_s}/\sigma_i \geq 2.5$, where $\sigma_i$ is the internal scatter of the time-series.}
         \label{FigSvM}
\end{figure}

Amongst the VVV-SkZ\_pipeline's outputs is a table of calibrated magnitudes of individual $K_s$-band frames. Unfortunately, not all observations were obtained in the weather conditions required for the VVV survey and about half of them had to be rejected. To verify our method for calibrating the $K_s$-band photometry in each individual frame, we verified that all frames share the same magnitude zero-point value by subtracting the magnitude of all observed stars with 12.5~mag~$< K_s < 14.5$~mag. For all frames, we get offsets close to 0, with an error comparable to the photometric accuracy. Figure\,\ref{FigSvM} shows the standard deviation, $\sigma_{K_s}$, as a function of the mean value, $K_s$, of the extracted light curves. A greyscale plot is used for better visibility. A theoretical relationship between $K_s$ and the photometric accuracy can be obtained starting with the definition of  magnitude:
\begin{equation}
  K_s=-2.5 \log{F I + R} + m_0,
\end{equation}
\noindent where $F$ is the mean value of the flat-field, $R$, the mean value of the readout, $m_0$, the magnitude zero-point and $I$ is the integrated observed total flux of a given star. Using the propagation of error formula, we obtain:
\begin{equation}
  \sigma_{K_s}=\frac{2.5 \log{e}}{I} \sqrt{(\sigma_F I)^2+I+\sigma_R^2},
  \label{sigKs}
\end{equation}
\noindent where $\sigma_F$ is the error on the flat-field (here, assumed to be 1\%) and $\sigma_R=\left[\sqrt{\pi} \,\rm{(size\,of\,the\,seeing)} \times \rm{(read noise\,per\,pixel)}\right]$ is the error on the readout. The only value that needs to be fitted to the data is $m_0$, and values of 23.8~mag and 23.5~mag were found for d084 and d086. This function is overplotted on the data in Figure\,\ref{FigSvM} with a solid green line. 

Due to saturation effects, the observed values of $\sigma_{K_s}$ deviate from Equation\,(\ref{sigKs}) when $K_s$ is \new{brighter} than $\sim$12.5~mag. Hence, for these magnitudes, we instead use a logarithmic function derived from fitting a slope to the $\log{\sigma_{K_s}}$ relation. Using this relation, one could determine a first list of candidate variables, setting a threshold at $5\times\sigma_{K_s}$. Stars meeting this criteria are marked with filled red circles in Figure\,\ref{FigSvM}. 

However, in addition to the intrinsic variability of stars, $\sigma_{K_s}$ contains systematic uncertainties that vary, depending on the position of the star on the detector (due to varying read noise and/or flat-field properties and/or presence of varying local background due to a nearby bright/saturated star). Hence, we adopt the approach described in D\'ekany et al. (\cite{De11}). During each acquisition of VVV data for fields d084 and d086, we systematically obtain for each epoch two frames within a few minutes. This allows us to get the internal scatter of the time-series, $\sigma_i$, defined as:
\begin{equation}
  \sigma_i=\sum_{j=1}^{N-1}(m_j-m_{j+1})^2/2M,
\end{equation}
\noindent where $m_j$ is the magnitude of the $j$-th detection. The sum is evaluated for M pairs of detections, each obtained within less than 5 minutes of each other. Candidate variable stars with $\sigma_{K_s}/\sigma_i \geq 2.5$ are shown as blue dots in Figure\,\ref{FigSvM}.

Unfortunately, due to bad weather, light curves rarely have more than 15 data points of an appropriate accuracy. Therefore, periodograms are very noisy and do not show any clear signal. On the other hand, some long term and/or high amplitude variations can be identified. We add to the Online Material a list of variable candidates contained in the four clusters, along with a brief description of their variations. Also, for both VVV regions studied, i.e. d084 and d086, we find a fraction of 1.5\% and 2\% of variable sources, respectively, which is in a agreement with the results of D\'ekany et al. (\cite{De11}) and Pietrukowicz et al. (\cite{Pi11}).
 
\onltab{2}{
\begin{table*}
\caption{List of variable candidates in Danks\,1 and Danks\,2. Columns include RA (J2000), DEC (J2000), Mean ($K_s$) and $\log{\sigma_{K_s}}$.}           
\label{table:5}      
\centering                          
\begin{tabular}{ccccccccccc}        
\hline\hline
\noalign{\smallskip}
Catalogue & RA (J2000) & DEC (J2000) & Mean($K_s$) & $\log{\sigma_{K_s}}$ && Catalogue & RA (J2000) & DEC (J2000) & Mean($K_s$) & $\log{\sigma_{K_s}}$ \\
number &                       &                          &                          &                                         && number &                       &                          &                          &                                        \\
\noalign{\smallskip}
\hline
\noalign{\smallskip}

D1-0078 & 198.07936 & -62.706541 & 15.86 & -0.80 & & D1-3546 & 198.12989 & -62.701732 & 16.08 & -0.89 \\
D1-0093 & 198.07970 & -62.705462 & 15.20 & -1.25 & & D1-3548 & 198.13037 & -62.713030 & 14.07 & -1.08 \\
D1-0093 & 198.08006 & -62.705399 & 15.79 & -0.85 & & D1-3540 & 198.13098 & -62.697912 & 14.45 & -1.10 \\
D1-0128 & 198.08058 & -62.705065 & 15.86 & -0.90 & & D1-3542 & 198.13103 & -62.700520 & 13.72 & -1.21 \\
D1-0119 & 198.08075 & -62.705865 & 16.28 & -0.72 & & D1-3550 & 198.13485 & -62.690548 & 18.83 & -0.08 \\
D1-0112 & 198.08141 & -62.707093 & 16.96 & -0.25 & & D1-3546 & 198.13841 & -62.701339 & 11.53 & -1.44 \\
D1-0276 & 198.08377 & -62.696737 & 14.32 & -1.11 & & D1-3548 & 198.13906 & -62.711442 & 17.82 & -0.03 \\
D1-0353 & 198.08523 & -62.701102 & 16.22 & -0.91 & & D1-3544 & 198.13994 & -62.704730 & 14.52 & -1.16 \\
D1-0365 & 198.08536 & -62.693862 & 13.07 & -1.14 & & D1-3542 & 198.14063 & -62.700532 & 14.78 & -1.00 \\
D1-0379 & 198.08566 & -62.704213 & 14.23 & -0.95 & & D1-3538 & 198.14074 & -62.709018 & 17.69 & -0.24 \\
D1-0520 & 198.08913 & -62.710759 & 18.51 & -0.29 & & D1-3538 & 198.14216 & -62.709617 & 16.76 & -0.28 \\
D1-0605 & 198.09067 & -62.704842 & 16.60 & -0.74 & & D1-3548 & 198.14252 & -62.710469 & 17.49 & -0.10 \\
D1-0624 & 198.09105 & -62.699257 & 17.31 & -0.52 & & D1-3538 & 198.14332 & -62.709620 & 15.35 & -0.71 \\
D1-0629 & 198.09112 & -62.703998 & 13.27 & -1.72 & & D1-3538 & 198.14337 & -62.708983 & 16.79 & -0.38 \\
D1-0719 & 198.09152 & -62.707777 & 16.73 & -0.67 & & D1-3544 & 198.14476 & -62.705129 & 16.82 & -0.55 \\
D1-0792 & 198.09356 & -62.712184 & 16.35 & -0.59 & & D2-0007 & 198.19466 & -62.686308 & 15.91 & -0.84 \\
D1-0823 & 198.09395 & -62.707737 & 11.97 & -0.82 & & D2-0045 & 198.19639 & -62.682487 & 17.74 & -0.45 \\
D1-0840 & 198.09415 & -62.692784 & 14.85 & -1.07 & & D2-0061 & 198.19709 & -62.684371 & 14.04 & -1.11 \\
D1-0848 & 198.09432 & -62.709843 & 17.63 & -0.14 & & D2-0187 & 198.20026 & -62.678689 & 13.91 & -1.22 \\
D1-0871 & 198.09473 & -62.702318 & 12.60 & -1.09 & & D2-0266 & 198.20216 & -62.675668 & 17.65 & -0.37 \\
D1-1124 & 198.09827 & -62.697101 & 17.24 & -0.51 & & D2-0342 & 198.20354 & -62.679100 & 18.57 & -0.05 \\
D1-1125 & 198.09835 & -62.703466 & 14.83 & -1.05 & & D2-0383 & 198.20495 & -62.670688 & 17.25 & -0.26 \\
D1-1295 & 198.10051 & -62.703707 & 14.17 & -1.05 & & D2-0415 & 198.20523 & -62.672838 & 16.06 & -0.88 \\
D1-1328 & 198.10091 & -62.716417 & 16.51 & -0.94 & & D2-0658 & 198.20920 & -62.672609 & 17.60 & -0.24 \\
D1-1506 & 198.10326 & -62.696218 & 12.64 & -1.30 & & D2-0992 & 198.21392 & -62.666613 & 12.98 & -0.93 \\
D1-1808 & 198.10722 & -62.700764 & 13.88 & -1.08 & & D2-1063 & 198.21452 & -62.683190 & 17.99 & -0.38 \\
D1-1941 & 198.10898 & -62.700922 & 14.09 & -1.16 & & D2-1519 & 198.22076 & -62.677181 & 17.48 & -0.31 \\
D1-2062 & 198.11031 & -62.714601 & 17.02 & -0.50 & & D2-1597 & 198.22225 & -62.689244 & 14.71 & -1.20 \\
D1-2350 & 198.11371 & -62.696060 & 12.40 & -1.10 & & D2-1862 & 198.22601 & -62.682288 & 13.30 & -0.81 \\
D1-2413 & 198.11389 & -62.697167 & 17.80 & -0.11 & & D2-2029 & 198.22850 & -62.679058 & 14.63 & -1.18 \\
D1-2379 & 198.11411 & -62.717062 & 15.75 & -1.04 & & D2-2182 & 198.23048 & -62.682038 & 15.86 & -1.09 \\
D1-2397 & 198.11427 & -62.687730 & 16.35 & -0.86 & & D2-2223 & 198.23107 & -62.674319 & 13.55 & -1.28 \\
D1-2408 & 198.11439 & -62.688650 & 12.16 & -1.60 & & D2-2299 & 198.23172 & -62.677520 & 17.72 & -0.37 \\
D1-2468 & 198.11514 & -62.712483 & 15.87 & -0.92 & & D2-2289 & 198.23204 & -62.694124 & 15.18 & -1.00 \\
D1-2763 & 198.11776 & -62.708380 & 18.45 & -0.32 & & D2-2497 & 198.23481 & -62.677910 & 15.51 & -0.84 \\
D1-2801 & 198.11893 & -62.708093 & 13.48 & -1.51 & & D2-2528 & 198.23522 & -62.670558 & 17.24 & -0.28 \\
D1-2879 & 198.11967 & -62.699463 & 14.46 & -1.27 & & D2-2548 & 198.23541 & -62.671228 & 16.95 & -0.29 \\
D1-2974 & 198.12079 & -62.692725 & 15.02 & -0.77 & & D2-2579 & 198.23612 & -62.694739 & 17.22 & -0.53 \\
D1-2974 & 198.12110 & -62.692668 & 15.38 & -1.13 & & D2-2599 & 198.23641 & -62.683018 & 14.71 & -1.39 \\
D1-3016 & 198.12155 & -62.707313 & 18.52 & -0.14 & & D2-2603 & 198.23733 & -62.690218 & 12.91 & -1.55 \\
D1-3135 & 198.12340 & -62.692690 & 17.57 & -0.17 & & D2-2783 & 198.23950 & -62.675108 & 12.78 & -1.49 \\
D1-3361 & 198.12551 & -62.701561 & 17.10 & -0.36 & & D2-2824 & 198.24012 & -62.667779 & 16.29 & -0.62 \\
D1-3465 & 198.12752 & -62.697527 & 13.99 & -1.46 & & D2-3128 & 198.24442 & -62.694468 & 18.28 & -0.19 \\
D1-3491 & 198.12779 & -62.692769 & 16.12 & -0.60 & & D2-3203 & 198.25360 & -62.692373 & 16.44 & -0.46 \\
D1-3554 & 198.12955 & -62.702017 & 17.23 & -0.18 & & & & & & \\
\noalign{\smallskip}
\hline
\end{tabular}
\end{table*}

\begin{table*}
\caption{List of variable candidates in RCW\,79. Columns include RA (J2000), DEC (J2000), Mean ($K_s$) and $\log{\sigma_{K_s}}$.}           
\label{table:6}      
\centering                          
\begin{tabular}{ccccccccccc}        
\hline\hline
\noalign{\smallskip}
Catalogue & RA (J2000) & DEC (J2000) & Mean($K_s$) & $\log{\sigma_{K_s}}$ && Catalogue & RA (J2000) & DEC (J2000) & Mean($K_s$) & $\log{\sigma_{K_s}}$ \\
number &                       &                          &                          &                                         && number &                       &                          &                          &                                        \\
\noalign{\smallskip}
\hline
\noalign{\smallskip}

RCW79-0029 & 204.94346 & -61.753372 & 14.71 & -1.42 & & RCW79-0445 & 204.98517 & -61.742081 & 18.54 & -0.21 \\
RCW79-0040 & 204.94577 & -61.759210 & 18.17 & -0.13 & & RCW79-0477 & 204.98628 & -61.750475 & 15.93 & -1.08 \\
RCW79-0002 & 204.94651 & -61.744302 & 18.04 & -0.42 & & RCW79-0520 & 204.98716 & -61.744896 & 17.09 & -0.27 \\
RCW79-0002 & 204.95280 & -61.738925 & 17.48 & -0.25 & & RCW79-0548 & 204.98786 & -61.746405 & 17.16 & -0.36 \\
RCW79-0029 & 204.95483 & -61.745262 & 17.68 & -0.32 & & RCW79-0634 & 204.98923 & -61.744089 & 16.33 & -0.43 \\
RCW79-0002 & 204.95519 & -61.737823 & 17.77 & -0.53 & & RCW79-0641 & 204.98962 & -61.750225 & 15.85 & -0.81 \\
RCW79-0029 & 204.96003 & -61.746189 & 17.58 & -0.04 & & RCW79-0669 & 204.99144 & -61.750221 & 17.58 & -0.41 \\
RCW79-0029 & 204.96382 & -61.742056 & 15.94 & -0.68 & & RCW79-0748 & 204.99336 & -61.736630 & 18.04 & -0.14 \\
RCW79-0150 & 204.96390 & -61.755874 & 17.87 & -0.48 & & RCW79-0847 & 204.99354 & -61.748245 & 17.95 & -0.31 \\
RCW79-0002 & 204.97027 & -61.734321 & 16.02 & -1.00 & & RCW79-0886 & 204.99462 & -61.736643 & 18.34 & -0.17 \\
RCW79-0131 & 204.97211 & -61.747668 & 18.14 & -0.18 & & RCW79-1028 & 204.99670 & -61.752871 & 16.31 & -0.92 \\
RCW79-0150 & 204.97213 & -61.754339 & 17.03 & -0.72 & & RCW79-0982 & 204.99693 & -61.741871 & 16.89 & -0.17 \\
RCW79-0150 & 204.97408 & -61.754537 & 17.26 & -0.46 & & RCW79-1019 & 204.99724 & -61.756554 & 11.96 & -0.96 \\
RCW79-0071 & 204.97458 & -61.741244 & 18.09 & -0.27 & & RCW79-0991 & 204.99758 & -61.740530 & 15.62 & -0.99 \\
RCW79-0176 & 204.97811 & -61.742926 & 17.79 & -0.09 & & RCW79-1088 & 204.99809 & -61.741124 & 18.03 & -0.04 \\
RCW79-0235 & 204.97847 & -61.750456 & 18.17 & -0.18 & & RCW79-1056 & 204.99852 & -61.747515 & 17.90 & -0.14 \\
RCW79-0330 & 204.98300 & -61.737352 & 16.47 & -0.79 & & RCW79-1207 & 205.00134 & -61.750141 & 17.62 & -0.31 \\
RCW79-0394 & 204.98390 & -61.736090 & 16.39 & -0.85 & & RCW79-1399 & 205.00368 & -61.743272 & 17.41 & -0.14 \\

\noalign{\smallskip}
\hline
\end{tabular}
\end{table*}

}

\section{Discussion}\label{Discussion}

\subsection{The G305 star forming complex}
This study provides additional information about the G305 star forming complex. In addition to the diffuse population of massive stars mentioned in Davies et al. (\cite{Da11}, see also Shara et al. \cite{Sh09} and Mauerhan et al. \cite{Ma11}), our spectral observations reveal 12 early B stars with distances that are comparable with Danks\,1 and Danks\,2. Therefore, although not cluster members they are definitely members of the G305 star forming complex. At this stage, it is hard to confirm if they are or are not runaway stars, because we cannot yet measure the proper motions using VVV data. Alternatively, they could be part of a larger association of young stars surrounding Danks\,1 and 2, formed within the same molecular cloud. 

\begin{figure}
  \centering
   \includegraphics[width=8cm]{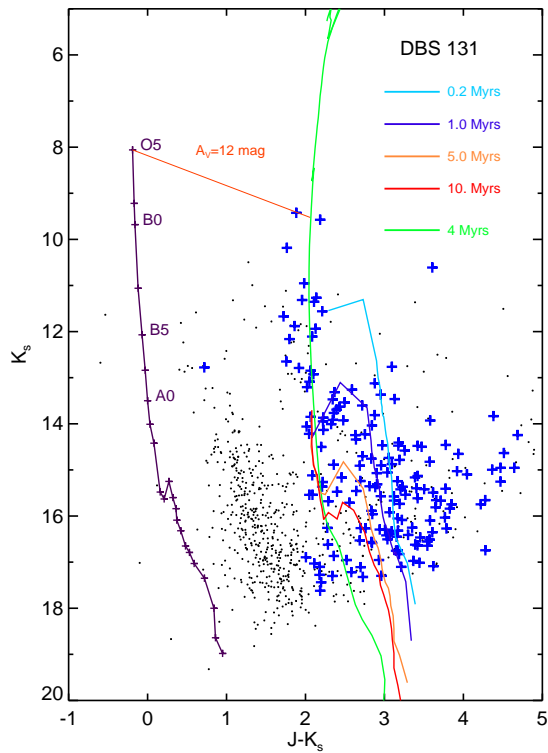}
      \caption{$(J-K_s)$ vs. $K_s$ diagram for DBS\,131, presented as in Figure\,\ref{FigCMD}. Dark dots are field stars and blue crosses are cluster stars.}
         \label{Fig131}
\end{figure}

Taking advantage of the VVV wide field of view, we examined the stellar content of the regions outlined by Hindson et al. (\cite{Hi10}). There are a number of already catalogued clusters in this region, namely: Danks\,1, Danks\,2, DBS\,83, DBS\,84, DBS\, 130, DBS\,131 (IR cluster G305.24+0.204, Clark et al. \cite{Cl04},  Leistra et al. \cite{Le05}, Longmore et al. \cite{Lo07}), DBS\,132, DBS\,133, DBS\,134, and G305.363+0.179 (Clark et al \cite{Cl04}). In addition to these, we have found three new young star clusters and/or stellar groups: VVV CL023, VVV CL022, VVV CL021 (Borissova et al. \cite{Bo11}), and a new star forming region SFR1.  The preliminary VVV CMD of G305.24+0.204 (DBS\,131) is shown in Figure\,\ref{Fig131}, where, in addition to a well defined MS, highlighted previously by Leistra et al. (\cite{Le05}) and  Longmore et al. (\cite{Lo07}), the PMS population is readily apparent in the VVV data. The adopted parameters of the plot are $m-M=12.85$~mag (3.72~ kpc), $E(J-K)=2.25$~mag, age 3--5~Myr. DBS\,131, like all other clusters from the G305 complex listed here, is very young (less than 5~Myr) and less massive than Danks\,1 and Danks\,2. A more detailed analysis of these objects will be presented in the next paper in our series (Borissova et al., in preparation). 

\subsection{Milky Way's structure}

The clusters' context within the Milky Way's structure is now assessed. The depth of the VVV photometry permits \new{the mapping} of crowded low latitude Galactic fields (Minniti et al. \cite{Mi11}). 
The youth of the stellar constituents inferred from the spectral types (Table\,\ref{table:2}) implies that Danks\,1 and Danks\,2 are viable tracers of the Galaxy's spiral structure. Georgelin et al. (\cite{Ge88}) first mentioned that the G305 region should be within the Scutum-Crux arm of the Galaxy. Later, Baume et al. \cite{Ba09} noted that Danks\,1 and Danks\,2 may belong to the Carina spiral arm. That conclusion is tied to their distance, which is significantly nearer ($>50$\%) than that found here. Their optical distance is acutely sensitive to variations in $R_V$, particularly since the total extinction in the optical $A_V=R_V \times E(B-V)$ is sizeable\new{, and not constrained by any spectroscopic observation. The present results} support the larger distance cited by Davies et al. (\cite{Da11}). 

\begin{figure}
  \centering
   \includegraphics[width=9.0cm]{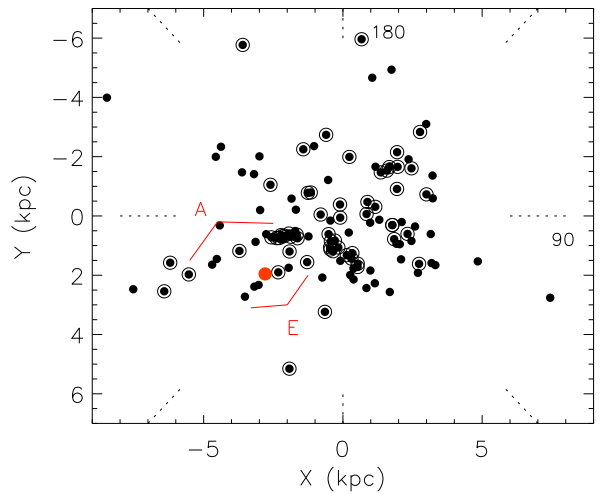}
      \caption{Map of local spiral structure as delineated by long period classical Cepheids (dots) and young clusters (circled dots) (see also Majaess et al. \cite{Maj09}). The Carina (A) and Centaurus (E) spiral arms are indicated on the diagram. Danks\,1 and Danks\,2 (red dot) reside in the Centaurus spiral arm.}
         \label{FigGS}
\end{figure}

The positions of Danks\,1 and Danks\,2 are plotted, in Figure\,\ref{FigGS}, on a hybrid map of the Galaxy's spiral structure, as delineated by long-period classical Cepheids and young ($<10$~Myr) open clusters. Long-period Cepheids are more massive younger stars than their shorter period counterparts, since the variables follow a period-luminosity relation. Cepheid variables and young open clusters define an analogous (local) spiral pattern (e.g. Majaess et al. \cite{Maj09}). Danks\,1 and Danks\,2 lie beyond the Sagittarius-Carina spiral arm and occupy the Centaurus arm, along with numerous young Cepheids and clusters (e.g., TW Nor, VW Cen, and VVV CL070). VVV CL070 was discovered in the comprehensive survey by Borissova et al. (\cite{Bo11}), who discovered 96 new clusters in the region sampled by the VVV survey.

\section{Summary}\label{Summary}

\new{In this paper we study three young known massive star clusters using data from the VVV survey and complementary low-resolution near-IR spectroscopy. 
Two of these clusters, Danks\,1 and Danks\,2, were investigated by Davies et al. (\cite{Da11}), whilst Martins et al. \cite{Ma10} provide excellent spectroscopic observations of the brightest members of RCW\,79. We choose these clusters to test and describe our methodology, which we will employ to study hundreds of open clusters observed by the VVV survey.  

The new analysis of Danks\,1 and Danks\,2 can be compared with that of Davies et al. (\cite{Da11}). We obtain comparable values of reddening, however, our results are based on many more spectra and use VVV $JHK_s$ CMDs and CCDs.  Our results account for differential reddening and the reddening laws were established for the line of sight, thus preventing uncertainties from propagating into the distance determination. Secondly, we also obtain similar distances. But, while their analysis was limited by the uncertainty on the luminosity class assignment, since they mostly observed bright stars that are near the turnoff, ours is less affected, as we have access to spectra of many B dwarf stars. The agreement with Davies et al. (\cite{Da11})'s results also calls into question the much smaller distances determined by Baume et al. (\cite{Ba09}).

We derive comparable cluster masses to those estimated by Davies et al. (\cite{Da11}). We also estimate that Danks\,1 and Danks\,2 \newII{are 3 and 5 Myr old respectively. Therefore, slightly older than Davies et al. (\cite{Da11}) suggest, but consistent within the uncertainties.} Taking advantage of the large field of view of VVV, we report a significant population of pre-main sequence stars. Interestingly enough, Danks\,2 contains the highest fraction (16$ \% $ of the total mass ) of PMS stars, followed by RCW\,79 ( 15$ \% $ ) and Danks\,1 (only  4$ \% $).  Based on radial velocity analysis, we found that Danks\,1 and Danks\,2 have the same mean radial velocity, which indicates that they are indeed binary clusters. A list of candidate variable stars is presented, which we will confirm as we obtain more VVV images. 

We also report our results for DBS\,132 and DBS\,131, open clusters situated near the two Danks clusters. Both clusters are very young (1-3 Myr), still embedded in dust and gas, and less massive than the Danks clusters. In addition to the diffuse population of massive WR stars, our spectral observations reveal 12 early B stars with distances that are comparable with Danks\,1 and Danks\,2. Therefore, although not cluster members they are members of the G305 star forming complex. The G305 complex most probably lies in the Centaurus arm, beyond the Sagittarius-Carina spiral arm. Finally, we present the first deep infrared colour-magnitude diagram of RCW79, revealing a large pre-main sequence population. We calculated the mass of this cluster to be roughly 3000 M$_{\odot}$.}

\begin{acknowledgements}

We would like to warmly acknowledge J.~P. Emerson for very fruitful discussions. This project is supported by the Chilean Ministry for the Economy, Development, and Tourism's Programa Iniciativa Cient\'{i}fica Milenio through grant P07-021-F, awarded to The Milky Way Millennium Nucleus. ANC gratefully acknowledges support from the Chilean Centro de Astrof\'isica FONDAP No. 15010003 and the Chilean Centro de Excelencia en Astrof\'isica y Tecnolog\'ias Afines (CATA) and Comitee Mixto ESO-GOBIERNO DE CHILE. JB is supported by FONDECYT  No. 1120601 and by the Ministry for the Economy, Development, and Tourism's Programa Inicativa Cient\'{i}fica Milenio through grant P07-021-F, awarded to The Milky Way Millennium Nucleus. JRAC is also supported by FONDECYT Regulat \#1080086. RK is supported by Centro de Astrofsica de Valparaso and Proyecto DIUV23/2009. DG also acknowledges FONDAP and CATA funds. DM is supported by FONDECYT Regular \#1090213, by the FONDAP Center for Astrophysics \#15010003, by the BASAL CATA Center for Astrophysics and Associated Technologies PFB-06, and by the MILENIO Milky Way Millennium Nucleus from the Ministry of Economy's ICM grant P07-021-F. RdG acknowledges partial research support through grant 11073001 from the National Natural Science Foundation of China. MSNK is supported by a Ci\^encia 2007 contract, funded by FCT/MCTES (Portugal) and POPH/FSE (EC).
            The data used in this paper have been obtained with NTT/SofI at the ESO La Silla Observatory, and with Clay/MMIRS at Las Campanas Observatory. This research has made use of the SIMBAD database, operated at CDS, Strasbourg, France. We gratefully acknowledge use of data from the ESO Public Survey programme ID 179.B-2002 taken with the VISTA telescope, and data products from the Cambridge Astronomical Survey Unit.

\end{acknowledgements}


\begin{thebibliography}{}
   \bibitem[2009]{Ba09} Baume, G., Carraro, G., \& Momany, Y.\ 2009, \mnras, 398, 221 
   \bibitem[1998]{Be98} Bessell, M.S., Castelli, F., \& Plez, B. 1998, A\&A, 333, 231
   \bibitem[2004]{Bi04} Bica, E., Ortolani, S., Momany, Y., Dutra, C. M. \& Barbuy, B. 2004, \mnras, 352, 226
   \bibitem[2005]{Bi05} Bik, A., Kaper, L., Hanson, M. M. \& Smits, M. 2005, A\&A, 440, 121
   \bibitem[2010]{Bon10} Bonatto C. \& Bica E. 2010, A\&A, 516, 81
   \bibitem[2011]{Bon11} Bonatto, C., \& Bica, E.  2011, \mnras, 415, 313 
   \bibitem[2011]{Bo11} Borissova, J., Bonatto, C., Kurtev, R. et al. 2011, A\&A, 532A, 131
   \bibitem[2012]{Bo12} Borissova, J., Georgiev, L., Hanson,  M.~M., Clarke, J.~R.~A., Kurtev, R., Ivanov, V.~D., Hillier, D.~J., \& Penaloza, F. 2012, A\&A, sumbmitted
   \bibitem[2011]{Ca11} Catelan, M., Minniti, D., Lucas, P. W. et al. 2011, in Carnegie Observatories Astrophysics Series, Vol. 5, RR Lyrae Stars, Metal-Poor Stars, and the Galaxy, ed. A. McWilliam, 145
   \bibitem[2004]{Cl04} Clark, J.~S. \& Porter, J.~M. 2004, A\&A, 427, 839
   \bibitem[1989]{Ca89} Cardelli, J.~A., Clayton, G.~C., \& Mathis 1989, ApJ,  345, 245
   \bibitem[2006]{Cr06} Crowther, P., Hadfield, L., Clark, J.,~S. Negueruela, I. \& Vacca, W. 2006, MNRAS, 372, 1407
   \bibitem[1983]{Da83} Danks, A.~C., Dennefeld, M., Wamsteker, W., \& Shaver, P.~A. 1983, A\&A,118, 301
   \bibitem[2012]{Da11} Davies, B., Clark, J.~S.; Trombley, C. et al. 2012, MNRAS, 419, 1871
   \bibitem[2011]{De11} Dekany, I., Catelan, M., Minniti, D. \& the VVV Collaboration 2011, arXiv:1111.0909v2
   \bibitem[2002]{Di02} Dias, W.~S., Alessi, B.~S., Moitinho, A., \& L\'epine, J.~R.~D. 2002, A\&A, 389, 871
   \bibitem[2002]{Du02} Dutra, C.~M., Santiago B.~X. \& Bica E. 2002, A\&A, 383, 219
   \bibitem[2007]{El07} Elmegreen, B.~G. 2007, ApJ, 668, 1064
   \bibitem[2010]{Em10} Emerson, J.~P. \& Sutherland, W. 2010, The Messenger, 139, 2
   \bibitem[2012]{Ek12} Ekstr{\"o}m, S., Georgy, C., Eggenberger, P., et al.\ 2012, \aap, 537, A146 
   \bibitem[1997]{Fi97} Figer, D. F., McLean, I. S. \& Najarro, F., 1997, ApJ, 486, 420
   \bibitem[2004]{Fi04} Fischera, J., \& Dopita, M.~A. 2004, ApJ, 611, 919
   \bibitem[1986]{Ge86} Gehrels, N., 1986, ApJ, 303, 336
   \bibitem[1988]{Ge88} Georgelin, Y.~M., Boulesteix, J., Georgelin, Y.~P., Le Coarer, E., \& Marcelin, M. 1988, A\&A, 205, 95
   \bibitem[2010]{Gi10} Girardi, L., Williams, B., Gilbert, K., Rosenfield, P., \& Dalcanton, J. 2010, ApJ, 724, 1030
   \bibitem[1996]{Ha96} Hanson, M.~M., Conti, P. S. \& Rieke, M. J. 1996, ApJS, 107, 281
   \bibitem[1998]{Ha98} Hanson, M.~M., Rieke, M. J. \& Luhman, K. L. 1998, AJ, 116, 1915 
   \bibitem[2005]{Ha05} Hanson, M.~M., Kudritzki, R.-P., Kenworthy, M. A., Puls, J. \& Tokunaga, A. T. 2005, ApJS, 161, 154
   \bibitem[2010]{Ha10} Hanson, M.~M., Kurtev, R., Borissova, J., Georgiev, L., Ivanov, V.~D., Hillier, D. J. \& Minniti, D. 2010 A\&A, 516A, 35
   \bibitem[1998]{Hi98} Hillier, D.~J., \& Miller, D.~L. 1998, ApJ, 496, 407
   \bibitem[2010]{Hi10} Hindson, L., Thompson, M.~A., Urquhart, J.~S., Clark, J.~S., \& Davies, B. 2010, \mnras, 408, 1438
   \bibitem[2004]{Ir04} Irwin, M. J., Lewis, J., Hodgkin, S., et al. 2004, in Society of Photo-Optical Instrumentation Engineers (SPIE) Conference Series, ed. P. J. Quinn \& A. Bridger, Vol. 5493, 411
   \bibitem[2005]{Iv05} Ivanov V.~D., Kurtev, R., \& Borissova, J. 2005, A\&A, 442, 195
   \bibitem[1983]{Ko83} Koornneef, J. 1983, A\&A, 128, 84
   \bibitem[1966]{Ki66} King, I.~R. 1966, AJ, 71, 64
   \bibitem[2011]{Kr11} Kroupa, P., Weidner, C., Pflamm-Altenburg, J., Thies, I., Dabringhausen, J., Marks, M., \& Maschberger, T. 2011, arXiv, arXiv:1112.3340
   \bibitem[2003]{La03} Lada \& Lada 2003, ARA\&A, 41, 57
   \bibitem[2001]{Le01} Lejeune, A., Schaerer 2001, A\&A, 366, 538
   \bibitem[2005]{Le05} Leistra, A., Cotera, A.~S., Liebert, J., \& Burton, M. 2005, AJ, 130, 1719
   \bibitem[2009]{Li09} Liermann, A., Hamann, W.-R. \& Oskinova, L. M. 2009, A\&A, 494, 1137
   \bibitem[2012]{Li12} Liermann, A., Hamann, W.-R., \& Oskinova, L.~M.\ 2012, \aap, 540, A14 
   \bibitem[1991]{Li91} Livingston W. \& Wallace L. 1991, National Solar Observatory: ``An atlas of the solar spectrum in the infrared from 1850 to 9000 cm-1'' , Technical Report \#91-001)
   \bibitem[2007]{Lo07} Longmore, S.~N., Maercker, M., Ramstedt, S., \& Burton, M. G. 2007, \mnras, 380, 1497
   \bibitem[2011]{Lo11} Longmore, A. J., Kurtev, R., Lucas, P. W., Froebrich, D., de Grijs, R., Ivanov, V.~D.V.~D., Maccarone, T. J., Borissova, J. \& Ker, L. M. 2011, \mnras, 416, 465
   \bibitem[1996]{Ma96} Maiolino, Rieke \& Rieke 1996, AJ, 111, 537
   \bibitem[2009]{Maj09} Majaess, D.~J., Turner, D. G., \& Lane, D. J. 2009, \mnras, 398, 263
   \bibitem[2011]{Maj11} Majaess, D.~J., Turner, D., Moni Bidin, C. et al. 2011,ApJ, 741L, 27
   \bibitem[2012]{Maj12} Majaess, D.~J., Turner, D., Moni Bidin, C. et al. 2012, A\&A, 537L, 4
   \bibitem[2005]{Ma05} Martins, F., Schaerer, D., \& Hillier, D.~J. 2005, \aap, 436, 1049 
   \bibitem[2007]{Ma07} Martins, L. P. \& Coelho, P. 2007, MNRAS, 381, 1329
   \bibitem[2008]{Ma08} Martins, F., Hillier, D.~J., Paumard, T., et al.  2008, \aap, 478, 219 
   \bibitem[2010]{Ma10} Martins, F., Pomar\`es, M., Deharveng, L., Zavagno, A., \& Bouret, J.-C. 2010, A\&A, 510, 32
   \bibitem[2011]{Ma11} Mauerhan, J., Van Dyk, S. \& Morris, P. 2011, AJ, 142, 40
   \bibitem[2012]{Ma12} Mauro, F., Moni Bidin, C., Chen\'e, A.-N et al. 2012, PASP, submitted
   \bibitem[2007]{Mo07} Monnier, J.~D., Tuthill, P.~G., Danchi, W.~C., Murphy, N. \& Harries, T.~J. 2007, ApJ, 655, 1033
   \bibitem[2010]{Mi10} Minniti, D., Lucas, P.~W., Emerson, J. P., et al. 2010, New A, 15, 433
   \bibitem[2011]{Mi11} Minniti, D., Saito, R., Alonso-Garc'a, J., Lucas, P.~W., \& Hempel, M. 2011, ApJ, 733L, 43 
   \bibitem[1996]{Mo96} Morris, Patrick W., Eenens, P. R. J., Hanson, Margaret M., Conti, Peter S. \& Blum, R. D. 1996, ApJ, 470, 597
   \bibitem[1998]{Mo98} Moorwood, A.,  Cuby, J.-G., \& Lidman, C. 1998, The Messenger, 91, 9
   \bibitem[2011]{Ne11} Negueruela, I., Gonz\'alez-Fern\'andez, C., Marco, A., \& Clark, J.~S. 2011, A\&A, 528, A59
   \bibitem[2007]{Ne07} Negueruela, I., Marco, A., Israel, G.~L. \& Bernabeu, G. 2007, A\&A, 471, 485 
   \bibitem[2010]{Ro10} Roeser, S., Demleitner, M., \& Schilbach, E.\ 2010, \aj, 139, 2440 
   \bibitem[2011]{Pi11} Pietrukowicz, P., Minniti, D., Alonso-Garc\'ia, J., \& Hempel, M. 2012, A\&A, 537A, 116
   \bibitem[2010]{Po10} Portegies Zwart et al. 2010, ARA\&A, 48, 431
   \bibitem[2009]{Pr09} Price \& Bate 2009, MNRAS, 398, 33
   \bibitem[2012]{Ra12} Ram{\'{\i}}rez Alegr{\'{\i}}a , S., Mar{\'{\i}}n-Franch, A., \& Herrero, A. 2012, A\&A, 541A, 75
   \bibitem[2009]{Ra09} Rayner, J.~T., Cushing, M.~C., \& Vacca, W.~D.\ 2009, \apjs, 185, 289 
   \bibitem[1985]{Ri85} Rieke, G. H. \& Lebofsky, M. J. 1985, ApJ, 288, 618
   \bibitem[2011]{Sa11} Sana, H., James, G., \& Gosset, E. 2011, MNRAS, 416, 817
   \bibitem[2012]{Sa12} Saito, R.~K., Hempel, M., Minniti, D. et al. 2012, A\&A, 537A, 107
   \bibitem[1998]{Sc98} Schlegel, D.~J., Finkbeiner, D.~P. \&  Davis, M. 1998, \apj, 500, 525
   \bibitem[1982]{Sc82} Schmidt-Kaler, T., 1982, in Landolt-Borstein, New Series, Group VI, vol. 2, ed. K. Schaifers \& H.~H. Voigt (Berlin: Springer-Verlag),1
   \bibitem[2008]{Sc08} Schnurr, O., Moffat, A.~F.~J., St-Louis, N., Morrell, N.~I. \& Guerrero, M.~A. 2008, \mnras, 389, 806
   \bibitem[2009]{Sh09} Shara, M. M., Moffat, A.~F.~J., Gerke, J. et al. 2009, AJ, 138, 402
   \bibitem[2000]{Se00} Siess L., Dufour E., \& Forestini M. 2000, A\&A, 358, 593
   \bibitem[2009]{de09} de Silva, G.~M., Gibson, B.~K., Lattanzio, J., \& Asplund, M. 2009, A\&A, 500L, 25
   \bibitem[1994]{St94} Stetson, P. B. 1994, PASP, 106, 250
   \bibitem[2008]{St08} Strai\v zys, V., \& Lazauskait\.e, R. 2008, Baltic Astronomy, 17, 253
   \bibitem[2009]{St09} Strai\v zys, V., \& Lazauskait\.e, R. 2009, Baltic Astronomy, 18, 19
   \bibitem[1976]{Tu76} Turner, D.~G. 1976, AJ, 81, 1125
   \bibitem[1989]{Tu89} Turner, D.~G. 1989, AJ, 98, 2300
   \bibitem[2011]{Tu11} Turner, D.~G. 2011, RMxAA, 47, 127
   \bibitem[2001]{va11} van der Hucht, K.~A. 2001, NewAR, 45, 135
   \bibitem[2012]{Wi12} Williams, P.~M., van der Hucht, K.~A., van Wyk, F., Marang, F., Whitelock, P.~A., Bouchet, P. \& Setia Gunawan, D.~Y.~A. 2012, \mnras, in press
   \bibitem[2005]{Wi05} de Wit, W.~J., Testi, L., Palla, F., \& Zinnecker, H. 2005, A\&A, 437, 247
   \bibitem[2001]{Ya01} Yadav R.~K.~S. \& Sagar R. 2001, MNRAS, 328, 370
   \bibitem[2011]{Zh11} Zhekov, S.~A., GagnŽ, M., \& Skinner, S.~L. 2011, ApJ, 727L, 17
     
\end{thebibliography}
\end{document}